\documentclass[sigconf, nonacm]{acmart}

\AtBeginDocument{%
  }

\usepackage{hyperref}
\usepackage{tikz}
\usepackage{multirow}
\usepackage{diagbox}
\usepackage{makecell}
\usepackage{titlesec}
\usepackage{algorithm2e}
\usepackage{algorithmic}
\usepackage[capitalise]{cleveref}
\usepackage{subcaption}
\usepackage{colortbl}

\newcommand{\DeformerName}{Surf-Deformer}
\newcommand{\frameworkName}{CaliScalpel}
\newcommand{\baselineLong}{Logical Swap for Calibration}
\newcommand{\baseline}{LSC}

\newcommand{\policyTwoName}{MS}

\titlespacing*{\section}
{0pt}{2.5ex plus 0.2ex minus .2ex}{0.3ex plus .2ex}

\newcommand{\todoFX}[1]{{\color{red} FX: #1}}

\newcommand{\KE}[1]{{\color{red} KE: #1}}
\newcommand{\YS}[1]{{\color{orange} YS: #1}}
\newcommand{\dean}[1]{{\color{purple} DT: #1}}

\begin{document}

\title{\frameworkName: In-Situ and Fine-Grained Qubit Calibration Integrated with Surface Code Quantum Error Correction}


\author{
Xiang Fang\textsuperscript{1}, Keyi Yin\textsuperscript{1}, Yuchen Zhu\textsuperscript{1}, Jixuan Ruan\textsuperscript{1}, Dean Tullsen\textsuperscript{1}, Zhiding Liang\textsuperscript{2}, Andrew Sornborger\textsuperscript{3}, Ang Li\textsuperscript{4}, Travis Humble\textsuperscript{5}, Yufei Ding\textsuperscript{1}, Yunong Shi\textsuperscript{6}}

\affiliation{
    \vspace{0.5em}
    \textsuperscript{1}University of California, San Diego, CA, USA \\
    \textsuperscript{2}Rensselaer Polytechnic Institute, Troy, NY, USA \\
    \textsuperscript{3}Los Alamos National Laboratory, Los Alamos, NM, USA \\
    \textsuperscript{4}Pacific Northwest National Laboratory, Richland, WA, USA \\
    \textsuperscript{5}Oak Ridge National Laboratory, Oak Ridge, TN, USA \\
    \textsuperscript{6}AWS Quantum Technologies, New York, NY, USA
    \country{}
}

\begin{abstract}
    Quantum Error Correction (QEC) is a cornerstone of fault-tolerant, large-scale quantum computing. However, qubit error drift significantly degrades QEC performance over time, necessitating periodic calibration. Traditional calibration methods disrupt quantum states, requiring system downtime and making {\it in situ} calibration infeasible. We present~\frameworkName, an innovative framework for {\it in situ} calibration in surface codes. The core idea behind \frameworkName~is leveraging code deformation to isolate qubits undergoing calibration from logical patches. This allows calibration to proceed concurrently with computation, while code enlargement maintains error correction capabilities with minimal qubit overhead. Additionally, \frameworkName~incorporates optimized calibration schedules derived from detailed device characterization, effectively minimizing physical error rates. Our results show that \frameworkName~achieves concurrent calibration and computation with modest qubit overhead and negligible execution time impact, marking a significant step toward practical {\it in situ} calibration in surface-code-based quantum computing systems.
\end{abstract}




\maketitle

\section{Introduction}
\label{sec: introduction}


Quantum Error Correction (QEC)~\cite{nielsen2010quantum, gottesman1998theory, shor1996fault, lidar2013quantum} is essential to enable large-scale Fault-Tolerant Quantum Computing (FTQC)  suitable for practical applications~\cite{shor1999polynomial, cao2019quantum}. Among various QEC schemes, 
\emph{surface codes}~\cite{bravyi1998quantum, dennis2002topological, litinski2019game, beverland2022surface, fowler2012surface} have emerged as the leading solution and have been successfully implemented on various quantum hardware~\cite{acharya2024quantumerrorcorrectionsurface, google2023suppressing, zhao2022realization, Bluvstein2024}. Surface codes work by encoding \emph{logical qubits} into redundant physical qubits arranged in a 2D square or hexagonal lattice, effectively reducing the error rate of the logical qubits if the physical error rate remains below a certain \emph{threshold}~\cite{fowler2012surface}. With increasing code size, surface codes provide exponential suppression of logical errors. Recent experiments on superconducting devices~\cite{huang2020superconducting, bravyi2022future} demonstrated this error suppression for the first time~\cite{acharya2024quantumerrorcorrectionsurface}, underscoring the practical applicability of surface codes. 

\begin{figure}[t]
\vspace*{0.1cm}
    \centering
    \includegraphics[width=0.47\textwidth]{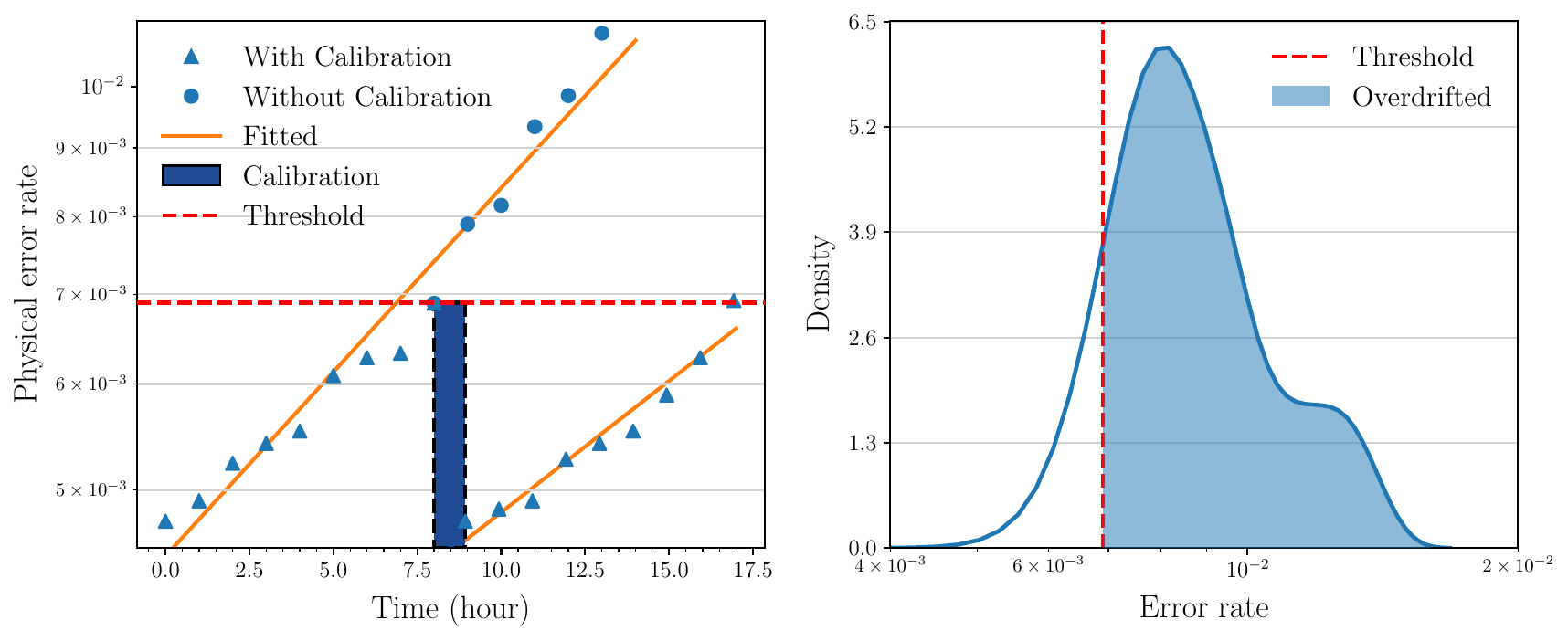} \\
    \hspace*{0.5cm} (a)  \hspace*{3.8cm} (b)    \vspace*{-0.3cm} \\
    \caption{Error drift on an IBM's device: (a) Calibration maintains low physical error rates. (b) Without calibration, a majority of qubits exceed the threshold after 24 hours.}
    \label{fig: error drift and calibration}
\vspace*{-0.5cm}
\end{figure}

Despite their demonstrated effectiveness, surface codes are highly sensitive to noise levels — small increases in physical error rates can require substantial expansion in code size to maintain the same logical error rate~\cite{SurfaceCode}. Consequently, maintaining a low and stable physical error rate is essential to ensure the effectiveness of surface codes during long computations. 
However, achieving such stability is challenging due to temporal variations in physical error rates in current quantum processors. Qubit and gate conditions degrade over time, leading to increased physical error rates—a phenomenon known as \emph{error drift}~\cite{proctor2020detecting, zhang2020error}. 
Fig.~\ref{fig: error drift and calibration}(a) illustrates the error drift observed on 
an IBM quantum computer~\cite{edman2024hardware}, showing that after just one day, over 90\% of single qubit gates exhibit error rates exceeding the threshold of surface codes. Theoretical estimates suggest that practical quantum applications are expected to run for hours or even days~\cite{Gidney2021howtofactorbit, femoco, litinski2023compute, babbush2018encoding}, assuming a fixed error rate. However, in practice, error drift can significantly undermine computational reliability, making such long computations infeasible.

\emph{Calibration}~\cite{kelly2018physical, klimov2020snake, tornow2022minimum, wittler2021integrated, tannu2019not} is the process of fine-tuning quantum hardware to maintain low physical error rates and counteract error drift. Quantum hardware installations frequently calibrate their devices-often several times a day-to keep qubits and gates aligned with optimal conditions. 
Calibration involves three key steps: characterization, control parameter adjustment, and validation. Importantly, the calibration process disrupts quantum states, making qubits under calibration unavailable for computation. 

The fundamental challenge in surface code based quantum computation lies at the intersection of two competing requirements: maintaining consistently low physical error rates (through calibration) while simultaneously preserving quantum information for ongoing computation. This tension motivates our central research question: {\it How can we integrate runtime calibration with ongoing computation while maintaining the protection level of surface codes?}


\begin{figure*}[!ht]
    \centering
    \includegraphics[width=.93\textwidth]{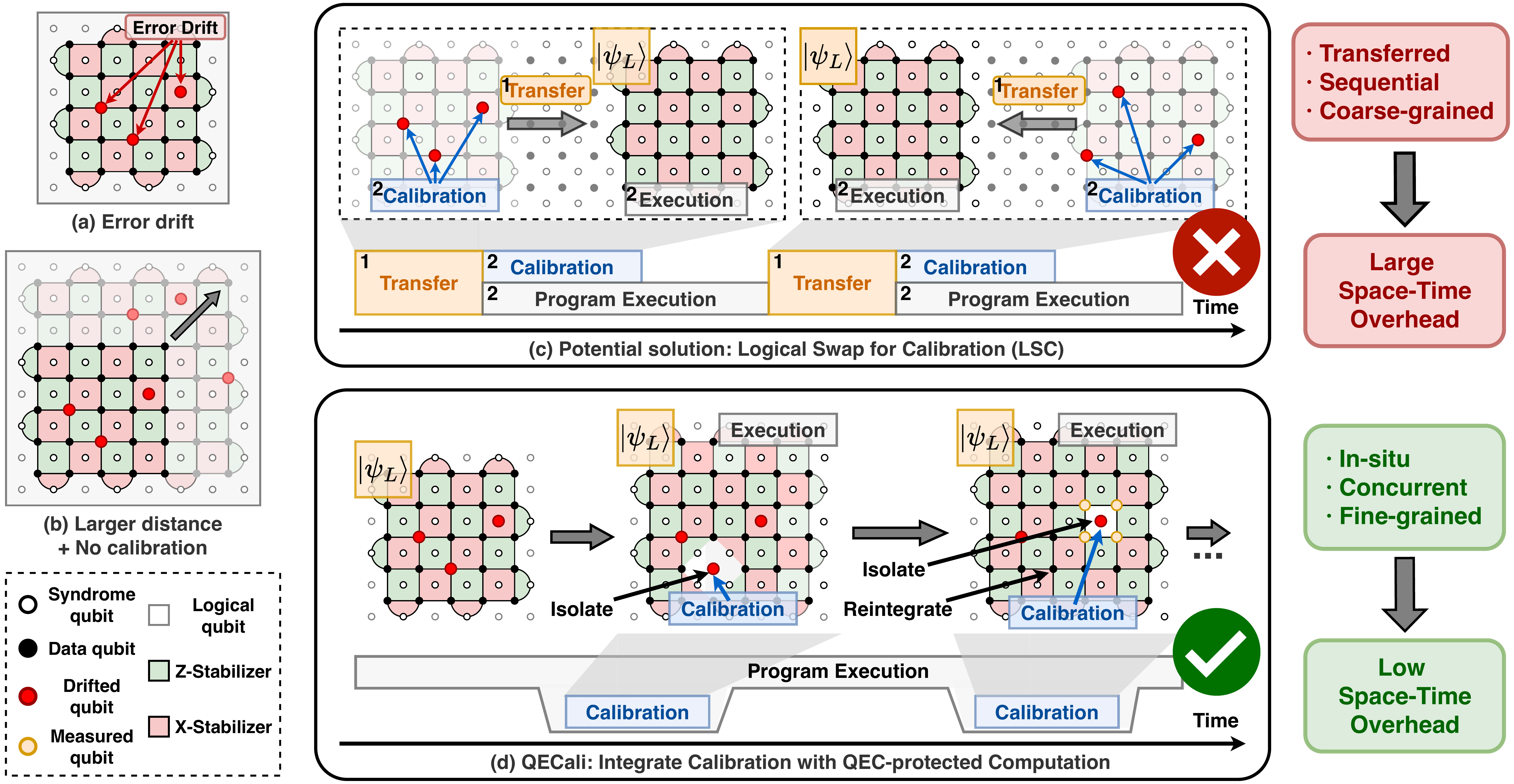}
    \caption{Comparison of~\frameworkName~and  \baseline. (a) Error drift pattern in physical qubits. (b) Increased code distance approach. (c)~\baseline, a coarse-grained approach with high overhead. (d)~\frameworkName~enables \textit{in situ} calibration.}
    \vspace*{-0.3cm}
    \label{fig: introduction}
\end{figure*}


A natural starting point draws inspiration from classical computing systems, where memory access must be coordinated with periodic DRAM refresh cycles [28]. Just as memory refresh temporarily interrupts access to maintain data integrity, qubit calibration requires periodic access to physical qubits while trying to preserve quantum information. 

However, key differences exist between the classical and quantum settings. While classical memory can be refreshed at the bit level, quantum error correction creates a critical constraint: individual physical qubits cannot be naively removed for calibration without catastrophic error propagation. Simply mirroring classical strategies leads to a conservative approach,  which we term \emph{\baselineLong} (\baseline): transferring entire logical qubits to ancillary regions via logical SWAP operations before calibrating their physical components~(\cref{fig: introduction}c). This classically inspired approach suffers from two fundamental limitations. First, 
 unlike classical bit copying, quantum logical SWAP operations require complex sequences of physical operations or quantum teleportation protocols~\cite{litinski2019game, beverland2022surface}, introducing substantial overhead in both ancilla qubit count and execution time. Second, there is a critical granularity mismatch—calibration operates at the physical qubit level, while computation occurs at the logical qubit level. Since physical qubits drift at varying rates~(\cref{fig: error drift and calibration}b),~\baseline~inefficiently forces entire logical qubits to pause when a single physical qubit requires calibration, leading to high overheads that scale with worst-case behavior.

To address these limitations, we propose \textbf{\frameworkName} (Fig.~\ref{fig: introduction}d), a novel QEC framework that enables \emph{in-situ} runtime calibration \emph{concurrent} with ongoing computation, while maintaining the same level of QEC protection provided by surface codes. This approach eliminates the substantial overhead associated with quantum state transfer and ancillary qubit preparation required by~\baseline. Furthermore, it operates in a \emph{fine-grained} manner, precisely isolating over-drifted physical qubits for calibration without stalling the entire logical qubit, effectively resolving the granularity mismatch issue of~\baseline. 

The advantages of~\frameworkName~are rooted in a key insight: {\it Surface codes allow for strategic runtime updates to their code structures, enabling qubit isolation while preserving logical information.} This is enabled by a theoretical tool known as {\it code deformation}~\cite{yin2024surf, vuillot2019code, bombin2009quantum}. Originally proposed for implementing logical operations on surface codes, we creatively repurpose it to address the conflicts between calibration and QEC-protected computation. Specifically, we apply the theory of code deformation to the most prevalent hardware architectures—square and hexagonal lattices~\cite{arute2019quantum, jurcevic2021demonstration, edman2024hardware}—and design novel code deformation instruction sets for these architectures (Sec.~\ref{sec: instruction set}). These instruction sets enable two complementary operations: \emph{selective qubit isolation}, which creates temporary boundaries to separate physical qubits for calibration, and \emph{dynamic code enlargement}, which slightly expands affected patches to maintain QEC capabilities during the calibration process. \frameworkName~combines this instruction set with intelligent scheduling (Sec.~\ref{sec: characterization},~\ref{sec: compiletime}) to create an efficient, automated calibration system while preserving the original QEC protection level.

Our evaluation on both simulation and real quantum hardware demonstrates QECali's effectiveness across different architectures and use cases. On practical quantum benchmarks such as quantum chemistry applications, QECali maintains sub-threshold error rates with only 12-15\% additional physical qubits and negligible impact on execution time, reducing retry risk by up to 85\% compared to baselines. Component-wise analysis shows that QECali's adaptive scheduling achieves 91\% reduction in calibration operations while minimizing space-time overhead. Our experiments on Rigetti and IBM processors validate QECali's key insight, showing that code deformation can effectively control error propagation.

In summary, this paper makes the following contributions:
\vspace{-3pt}
\begin{itemize}
\item We propose a novel framework, \textbf{\frameworkName}, that enables in-situ calibration and concurrent QEC-protected computation, achieving the desired retry risk level with minimal qubit count overhead and negligible execution time overhead.

\item We design and formalize novel surface code deformation instruction sets for calibration on square and heavy-hexagon topologies supporting the \frameworkName~framework. 

\item We develop an adaptive scheduling method to efficiently coordinate calibration operations across physical qubits while effectively balancing various constraints.


\item We introduce key device metrics—calibration times, drift rates, and crosstalk—to characterize quantum hardware, providing critical insights for optimizing calibration schedules and predicting performance.

\end{itemize}

\section{Background}
\label{sec:background}
This section provides an overview of surface codes and their deformation operations, and background on error drift and calibration.

\subsection{QEC with Surface Code}
\begin{figure}[!ht]
\centering
\includegraphics[width=0.47\textwidth]{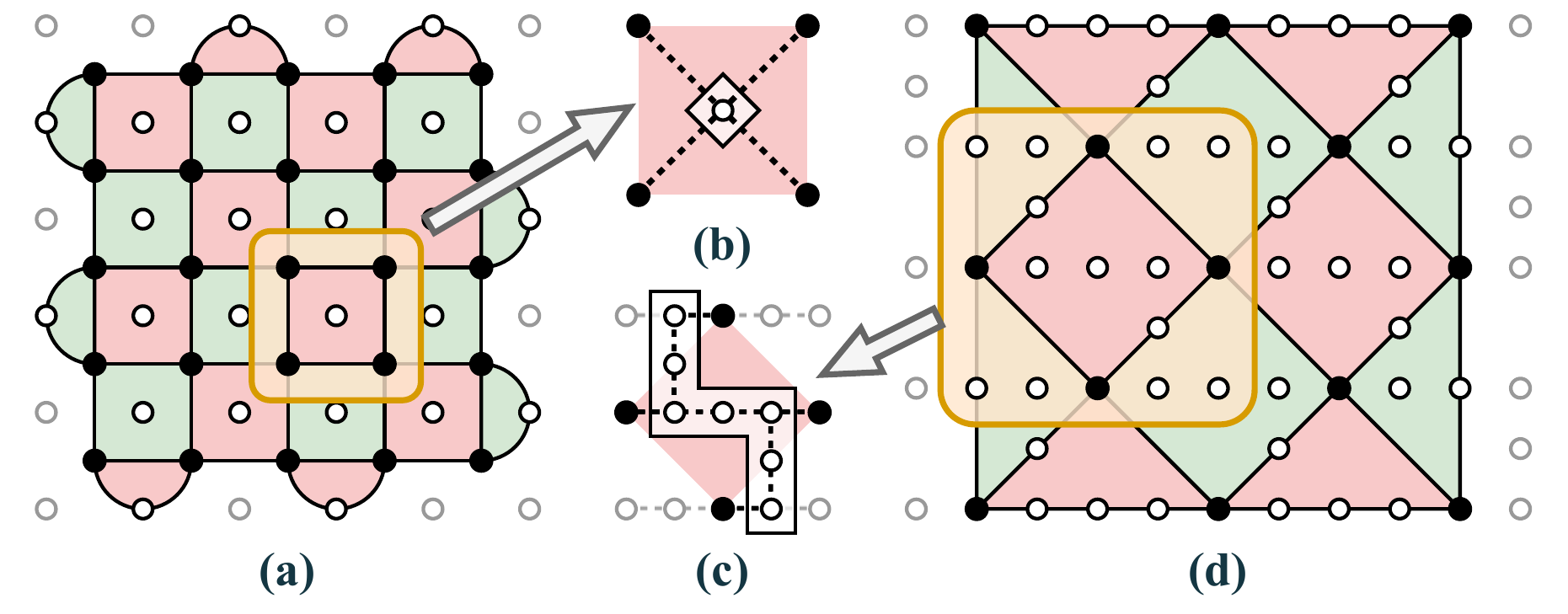}
\caption{(a)(d) Surface codes on a square and hexagon lattice. (b)(c) Stabilizers of surface codes on two lattices.}
\label{fig background}
\end{figure}

{\label{subsec: surface code}}
\noindent\textbf{Surface Code Basics. }The surface code is a leading candidate for realizing FTQC due to its simple 2D grid structure and high error threshold ($1\%$)~\cite{bravyi1998quantum, fowler2012surface}.~\cref{fig background}a presents a typical surface code patch, where multiple physical qubits encode a single \emph{logical qubit}. The physical qubits are divided into \emph{data qubits} (black dots), which store the logical quantum state, and \emph{syndrome qubits} (white dots), which collect error information from neighboring data qubits. The surface code can also be implemented on a hexagonal topology, as shown in Fig.\ref{fig background}b, which reflects the architecture of some current hardware platforms (IBM’s devices~\cite{edman2024hardware}). Unlike in the square lattice, where each stabilizer uses a single ancilla qubit as the syndrome qubit, the stabilizers in the heavy-hexagon structure use seven ancilla qubits arranged in an "S" shape to form a bridge connecting the four data qubits, as shown in \cref{fig background}(c). This "S"-shaped arrangement provides the connectivity needed to extract error syndromes while mitigating the frequency crowding issue.

\vspace{3pt}
\noindent\textbf{QEC with Surface Codes. }Each syndrome qubit is linked to a \emph{stabilizer}~\cite{gottesman1996class, gottesman1998heisenberg, nielsen2010quantum}, a Pauli operator involving adjacent data qubits, as shown in Fig.\ref{fig background}b,c. The red and green colors indicate the two stabilizer types, comprising exclusively $X$- or $Z$-Pauli operators, respectively. The syndrome qubits provide measurement outcomes of these stabilizers, producing \emph{error syndromes} to guide error correction. However, some errors can alter the logical state without being detected by stabilizer measurements, known as \emph{logical errors}, resulting in program failure. The \emph{logical error rate} (LER) is closely tied to the \emph{code distance}, defined as the minimum number of single-qubit errors required to cause a logical error. The code distance of a perfect surface code equals the number of data qubits along its edge (a distance-$5$ surface code in Fig.~\ref{fig background}a). The LER decays exponentially with increasing code distance, provided the physical error rate remains below a certain threshold. For further detail, see~\cite{fowler2012surface, dennis2002topological, kim2023design}.

\subsection{Deformation of Surface Codes}
\label{sec deformation}


Code deformation~\cite{yin2024surf, vuillot2019code, bombin2009quantum} is a key technique in surface codes that modifies the shape of the code patch to achieve specific functions. This subsection introduces four key deformation instructions that, when combined, isolate qubits from the code patch while preserving the encoded quantum information and QEC capability, enabling calibration without disrupting ongoing computation.

\begin{figure}[!ht]
\centering
\includegraphics[width=0.47\textwidth]{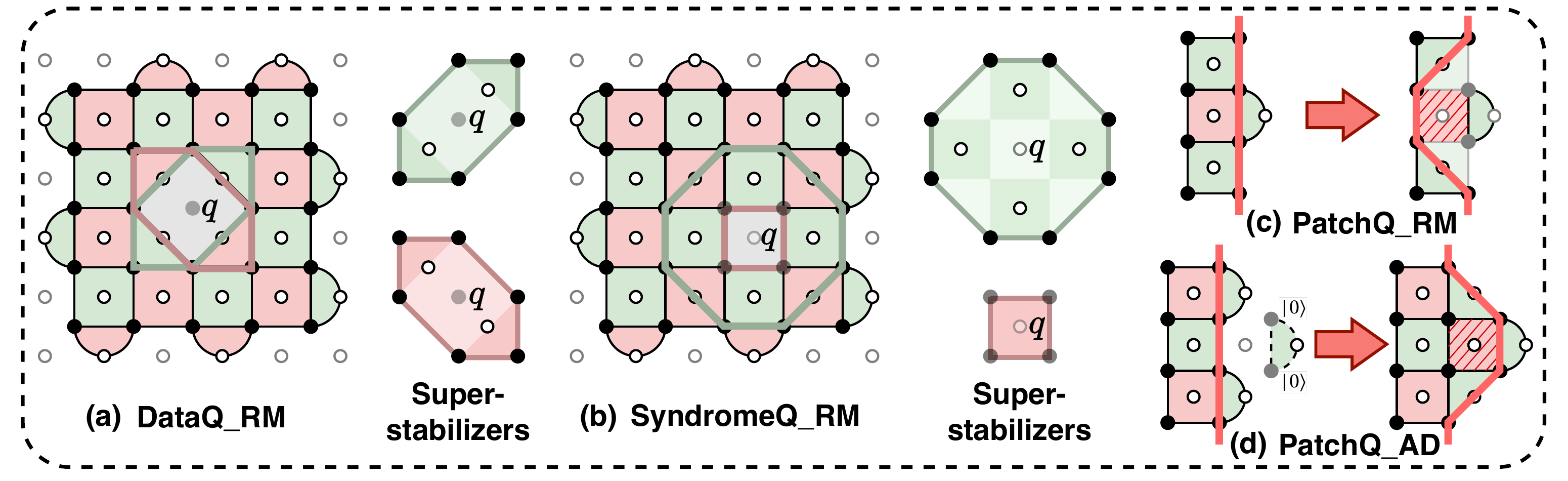}
\caption{Four key deformation operations. (a) Data qubit removal, (b) Syndrome qubit removal, (c) Patch shrinkage at the boundary, and (d) Patch expansion at the boundary.}
\label{fig deform}
\end{figure}

\noindent 1. \texttt{DataQ\_RM.} As illustrated in Fig.~\ref{fig deform}a, to remove the data qubit $q$, it combines the original stabilizers into larger “superstabilizers” that exclude $q$~\cite{stace2009thresholds, auger2017fault, stace2010error, nagayama2017surface}. These superstabilizers, along with the remaining stabilizers, enable QEC to continue on the code patch consisting of the remaining qubits during subsequent QEC cycles. To reintegrate the data qubit $q$ into the system, the stabilizers of the original code are measured, restoring the original structure.

\noindent 2. \texttt{SyndromeQ\_RM.} As shown in Fig.~\ref{fig deform}b, to remove the syndrome qubit $q$, the qubits involved in the stabilizer associated with $q$ are measured in the $X$- or $Z$-basis, depending on the stabilizer type. The stabilizers are then updated to form “superstabilizers” that exclude $q$ and are used in subsequent QEC cycles. Reintegrating the syndrome qubit into the system is accomplished by measuring the original stabilizers.

\noindent 3. \texttt{PatchQ\_RM.} This operation shrinks the code patch at the boundary by measuring the qubits to be excluded in either the $Z$- or $X$-basis, depending on the stabilizer they are associated with. To reintegrate the isolated qubits, the original stabilizer (the red one in Fig.~\ref{fig deform}c) is measured.

\noindent 4. \texttt{PatchQ\_AD.} This operation expands the code patch at the boundary by preparing the qubits to be included in appropriate states ($|0\rangle$ or $|+\rangle$) and measuring the new stabilizer. In the example shown in Fig.~\ref{fig deform}d, the qubits are initialized in the $|0\rangle$ state, and a new $X$-stabilizer (red) is measured.

By combining \texttt{DataQ\_RM}, \texttt{SyndromeQ\_RM}, and \texttt{PatchQ\_RM}, various patterns of drift-affected qubits can be isolated from the code patch. Repeated application of \texttt{PatchQ\_AD} enables the desired code enlargement, restoring QEC capability after qubit isolation. For a more comprehensive introduction to the deformation framework in surface codes, refer to~\cite{yin2024surf, vuillot2019code, bombin2009quantum}.

\begin{figure*}[!ht]
    \centering
    \includegraphics[width=1.0\textwidth]{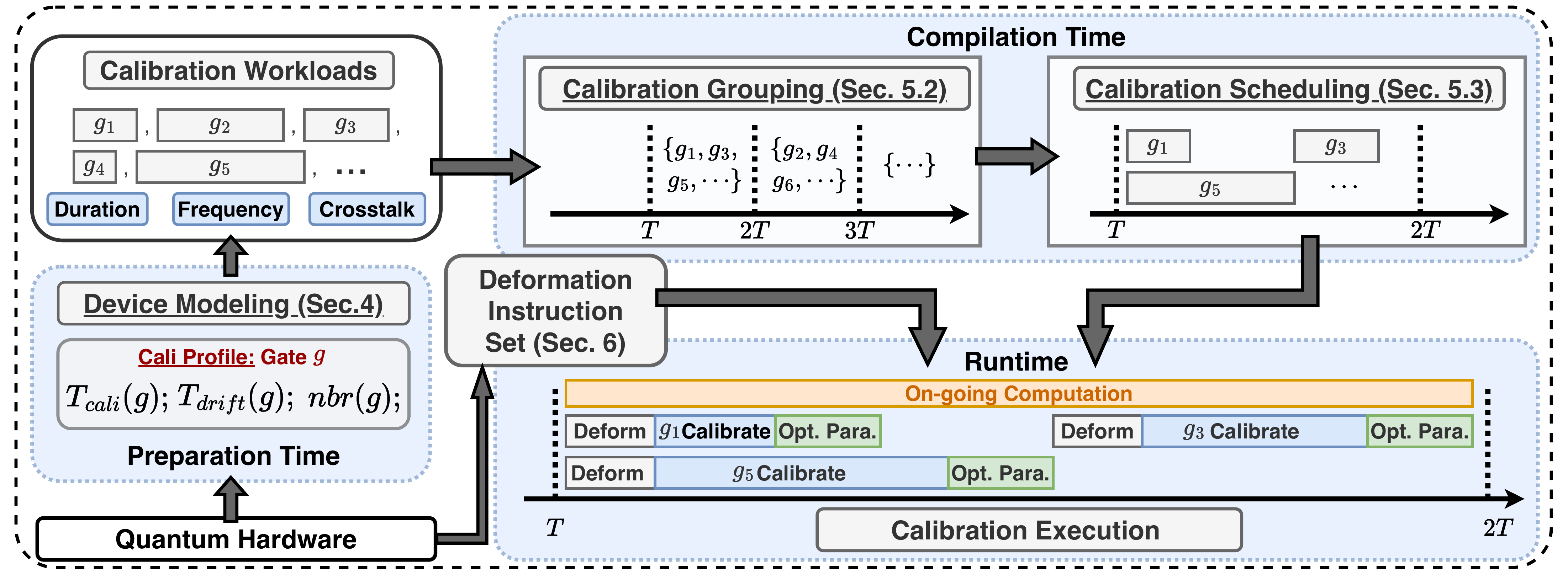}
    \caption{Overview of \frameworkName: a deformation-based calibration framework.}
    \label{fig: workflow}
\end{figure*}



\subsection{Error Drift and Calibration}
\noindent\textbf{Error Drift. }QEC schemes often assume a static error model where error rates remain constant. However, on real hardware, error rates fluctuate over time, potentially exceeding the QEC threshold—a phenomenon known as \emph{error drift}~\cite{ravi2023navigating, proctor2020detecting, zhang2020error}. In superconducting devices, a primary cause of error drift is the unwanted coupling of qubits to \emph{two-level systems} (TLS)~\cite{martinis2005decoherence, muller2019towards, klimov2018fluctuations}, along with thermal fluctuations in Josephson junctions~\cite{gumucs2023calorimetry}, unpaired electrons~\cite{gustavsson2016suppressing, klimov2018fluctuations}, and environmental noise~\cite{martinis2021saving, mcewen2022resolving}. TLS defects, inherent to the fabrication process, occur randomly and are difficult to characterize or predict, leading to unique and heterogeneous impacts on each qubit. These factors cause qubit parameters (e.g., $T_1$ and $T_2$ coherence times) to vary unpredictably, necessitating frequent characterization and recalibration to maintain optimal performance.

\vspace{3pt}
\noindent\textbf{Calibration. } \emph{Calibration}~\cite{neill2018blueprint, kelly2018physical, arute2019quantum, liu2023enabling, tornow2022minimum, wittler2021integrated, klimov2024optimizing, li2024high} is the process of readjusting control parameters (such as the duration and frequency of control pulses) for qubit operations to maintain optimal performance, specifically, low error rates. 
This process is lengthy, often taking several hours (e.g., 4 hours in~\cite{arute2019quantum}) to ensure accuracy. Moreover, the calibration order for different qubits must be carefully scheduled to mitigate crosstalk issues~\cite{murali2020software}. Importantly, calibration requires qubits to be in specific states for measurement, preventing them from storing information or supporting computation during calibration. As a result, calibration typically halts the program, making concurrent calibration and computation impossible and limiting the duration of reliable computations.


\section{Overview}
\label{sec:overview}
\frameworkName~achieves efficient in-situ calibration through three coordinated stages, as depicted in \cref{fig: workflow}:

\noindent\textbf{Preparation time. }This stage thoroughly characterizes the quantum hardware, capturing key parameters for each gate, including \emph{calibration duration}, \emph{drift rate}, and \emph{calibration crosstalk} (\cref{subsec: technical subsec1}). This foundational data serves as a basis for optimizing our calibration strategies. 

\noindent\textbf{Compilation time.} This stage aims to determine all the calibration processes based on the hardware characterization obtained during preparation stage. This stage consists of two steps.

\noindent\emph{Drift-based Calibration Grouping (Sec.~\ref{subsec: technical subsec2}).} Based on gate parameters, we define calibration workloads by determining each gate’s frequency and duration. Gates with similar drift characteristics are grouped into calibration intervals to avoid overlapping workloads, which could cause excessive distance loss from deformation.

\noindent\emph{Intra-Group Calibration Scheduling (Sec.~\ref{subsec:technical-subsec3})}: Within each interval, we sequence the calibration workloads based on their dependencies and crosstalks. This scheduling minimizes the total calibration time while maintaining computational efficiency.

\noindent\textbf{Runtime. }
The runtime follows the calibration schedule generated during compile time (\cref{sec: compiletime}), triggering the corresponding calibration operations for designated gates at specified intervals. For each gate, it executes the associated code deformation instructions from the \frameworkName~instruction set (\cref{sec: instruction set}). Meanwhile, logical computations on deformed logical qubits continue uninterrupted, with all operations pre-calculated during compile time.




\section{Preparation-time Device Characterization}{\label{sec: characterization}}
\label{subsec: technical subsec1}
To enable optimized calibration scheduling, ~\frameworkName~first characterizes the quantum device pre-compilation by measuring and extracting key metrics of each qubit operations. These include:

\noindent\textbf{Calibration Time} ($T_{\text{cali}}$): We measure each gate's calibration duration through repeated experiments, as this time directly impacts scheduling efficiency. 

\vspace{3pt}
\noindent\textbf{Drift Rate} ($T_{\text{drift}}$): We characterize how quickly a gate $g$'s error rate doubles using the following formula,
\begin{equation}
\label{eq: Drift Constant Time}
       p(g, t) = p_0[g] \cdot 10^{t / T_{\text{drift}}[g]}
\end{equation}
where $p(g,t)$ represents the error rate at time $t$, $p_0[g]$ is the initial rate, and $T_{\text{drift}}[g]$ is the drift time constant. We determine these parameters by repeatedly measuring gate error rates.

\begin{figure}[!ht]
  \vspace*{-0.4cm}
    \centering
    \includegraphics[width=0.43\textwidth]{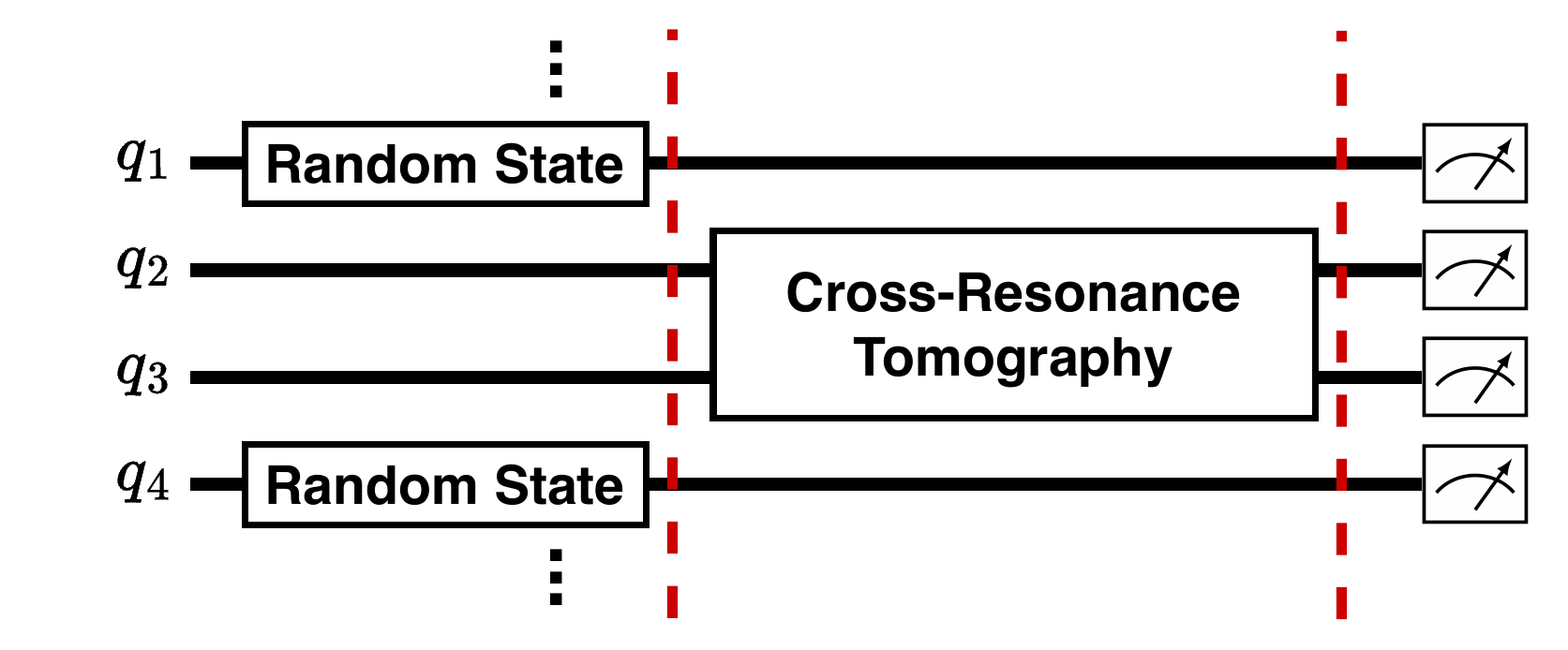}
    \caption{Circuit used for characterizing calibration-induced crosstalk.}
    \label{fig: cali procedure}
  \vspace*{-0.4cm}
\end{figure}

\vspace{3pt}
\noindent\textbf{Calibration Crosstalk} ($\text{nbr}(g)$): We identify qubits affected by each gate's calibration using the circuit in \cref{fig: cali procedure}. 
For each gate $g$, we initialize nearby qubits to unknown states, perform calibration, and measure final states. Qubits showing deviations beyond a threshold are added to $\text{nbr}(g)$, indicating crosstalk interference.

These metrics form the foundation for QECali's compilation-time scheduler (Section~\ref{sec: compiletime}) and runtime execution engine, enabling generation of calibration schedules that optimize the critical trade-offs between frequency, parallelism, and interference.

\section{Compilation-time Calibration Scheduling}
\label{sec: compiletime}
In this section, we delve into ~\frameworkName's compilation-time calibration scheduling. First, we group qubits with similar drift patterns to enable coordinated calibration, then optimize scheduling within these groups. This structured approach allows us to balance the competing demands of individual qubits while managing constraints like qubit resource limitations and crosstalk.

\subsection{Problem Formulation}
\label{subsec:problem}
For reliable execution of large-scale quantum programs, we must maintain the logical error rate (LER) below a target value (say $\text{LER}_{\text{tar}}$) throughout computation.  Formally, this requires \(\text{LER}(t) \leq \text{LER}_{\text{tar}}\) for all \(t\). The calibration schedule directly impacts this requirement by influencing the physical error rates: frequent calibration helps maintain lower physical error rates, which in turn supports a lower LER. 

For a distance-\(d\) surface code to achieve the target \(\text{LER}_{\text{tar}}\), the average physical error rate must not exceed a corresponding target, denoted as \(p_{\text{tar}}\). Thus, each gate g must be calibrated within its drift time $T_{\text{drift}, p_{\text{tar}}}[g]$—the time it takes for $g$'s error rate to reach $p_{\text{tar}}$. To meet the reliability requirement, the calibration interval for \(g\) must satisfy $T_g \leq T_{\text{drift},p_{\text{tar}}}[g]$. The drift time \(T_{\text{drift},p_{\text{tar}}}[g]\) can vary significantly depending on the specific gate and hardware characteristics, ranging from hours to days. 

While frequent calibration is necessary for maintaining low LER, parallel calibration comes with substantial costs. When calibrating $k^2$ physical qubits simultaneously in a $d\times d$ surface code patch, we can derive that $4kd + 4k^2$ physical qubits need to be added to maintain code distance $d$—a significant overhead for large $k$. Moreover, calibration-induced crosstalk between neighboring qubits further constrains the degree of parallelization possible.

To balance these competing requirements, we formulate our optimization objective as:
\begin{align*}
\min \sum_g \frac{1}{T_g}  \ \  \text{ subject to }\ \ 
\underbrace{T_g \le T_{\text{drift}, p_{\text{tar}}}[g],}_\text{drift constraint} \ \   \underbrace{|C_t| \leq 1, \forall t  }_{\text{crosstalk constraint}}
\end{align*}
 where $T_g$ is the calibration period for gate g, $C_t$ is the set of gates in $C$ being calibrated at time $t$, and $C$ represents a set of gates that cannot be calibrated simultaneously due to crosstalk. This formulation aims to minimize calibration frequency, for reducing qubit overhead, while respecting both drift time limits and crosstalk constraints.

\subsection{Drift-based Calibration Grouping}
\label{subsec: technical subsec2}

To solve this problem, we propose a heuristic solution that achieves high optimization quality with fast compilation speed. Our approach groups gates with similar drift characteristics to share calibration intervals, effectively discretizing the scheduling space while respecting device physics. Specifically, we select a base calibration interval $kT_\text{Cali}$ and assign each gate $g$ to a calibration group $k_g$ according to:
\begin{equation}
   \label{eq:grouping}
   k_g \cdot T_\text{Cali} \leq T_{\text{drift}, p_{\text{tar}}}[g] < (k_g + 1) \cdot T_\text{Cali}
\end{equation}

This grouping strategy offers several advantages. First, gates within group $k_g$ share the same calibration cycle $T_g = k_g \cdot T_\text{Cali}$, simplifying scheduling. 
Once the grouping is completed, the schedule for each group is determined: during the \(k\)-th interval, only gates belonging to Group \(k_g\) are executed (where \(k \mod k_g = 0\)).  
Second, this grouping enables opportunities for parallel scheduling of calibrations, as all gates within a group can be scheduled at any time within their assigned interval \(T_\text{Cali}\).  
Third, it simplifies the issue of crosstalk, as we only need to address crosstalk within a single group. Crosstalk between groups is avoided since groups assigned to the same interval are trivially scheduled sequentially.  The resulting calibration frequency becomes:
\begin{equation}
   \label{eq:frequency}
   \sum_g \frac{1}{T_g} = \frac{1}{T_\text{Cali}}\sum_{k} \frac{n_k}{k}
\end{equation}
where $n_k$ denotes the number of gates in the k-th group.

\begin{figure}[t]
    \centering
    \includegraphics[width=0.47\textwidth]{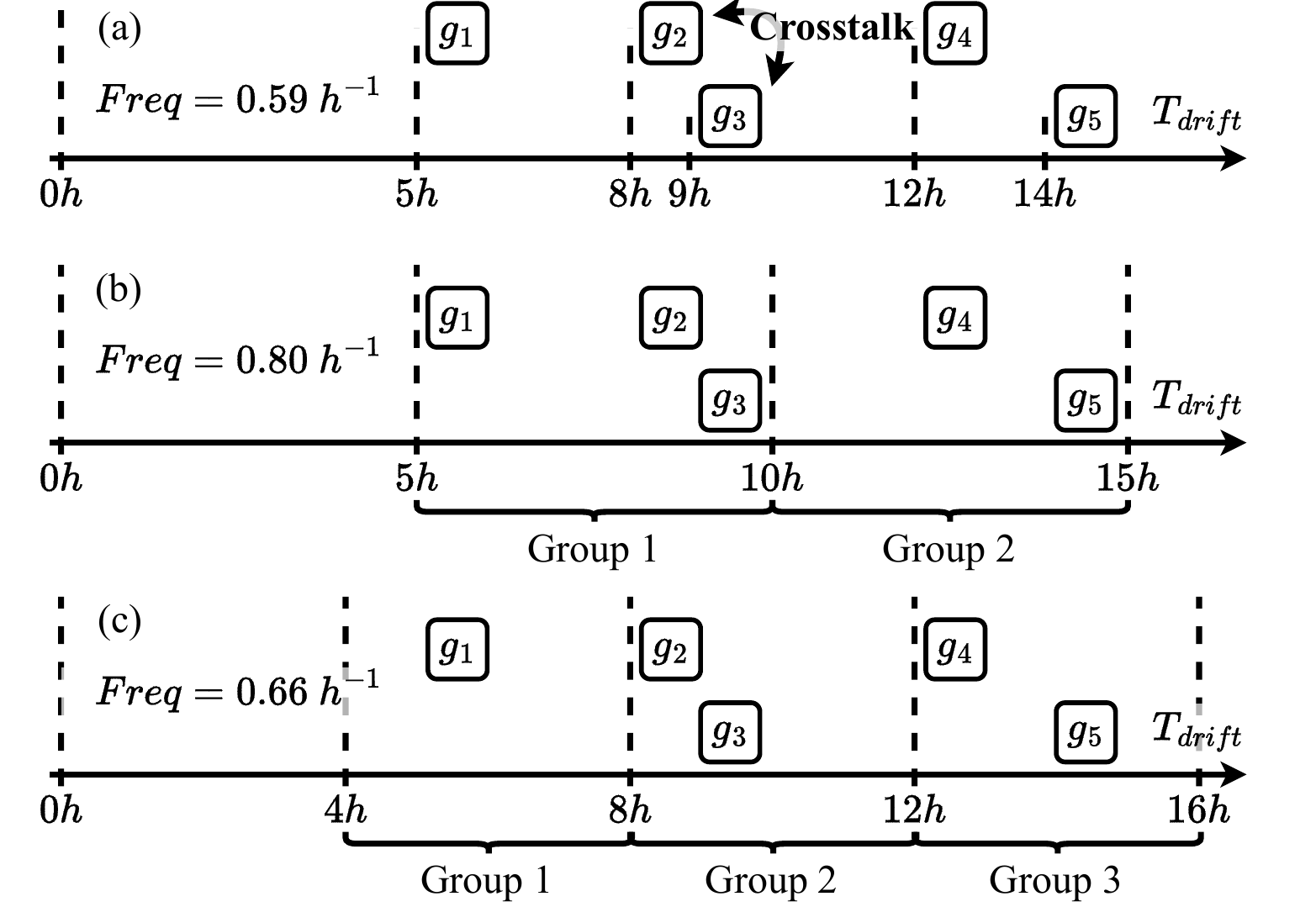}
    \caption{Impact of \(T_\text{Cali}\) on calibration frequency.}
    \label{fig: Adaptive calibration assigning}
\end{figure}


\noindent\textbf{Optimal Choice of Group Duration \(T_\text{Cali}\):}  
The choice of base calibration interval $T_\text{Cali}$ significantly impacts grouping efficiency. While one might intuitively set \(T_\text{Cali}\) to the minimum drift time \(\min_g T_{\text{drift},p_{\text{tar}}}[g]\),  this often leads to suboptimal groupings. Consider the example in  \cref{fig: Adaptive calibration assigning}(a): with \(T_\text{Cali} = 5h\), gates \(g_1\), \(g_2\), and \(g_3\) form Group 1 while gates \(g_4\) and \(g_5\) form Group 2, resulting in  0.80 calibrations per hour.  
However, setting \(T_\text{Cali} = 4h\), though increasing  \(g_1\)'s calibration frequency, enables better distribution of other gates into groups with lower frequencies, reducing overall cost to 0.66 calibrations per hour. 

\RestyleAlgo{ruled}
\LinesNumbered
\begin{algorithm}[hbt!]
\caption{Calibration Group Assignment}
\label{alg: grouping}
\KwIn{
Gate set $G$, Drift time $T_{\text{drift}, p_{\text{tar}}}[g]$
}
\KwOut{
Calibration groups $Group[k]$
}
$T_\text{min} = \min_g T_{\text{drift}, p_{\text{tar}}}[g]$ \\
$T_\text{Cali} = T_\text{min}$ \\
\For{$g \in G$}{
    $k = \left\lceil{T_{\text{drift}, p_\text{tar}}[g] / T_\text{min}}\right\rceil$ \\
    $T = T_{\text{drift}, p_\text{tar}}[g] / k$ \\
    \If{$Freq(G, T_\text{Cali}) > Freq(G, T)$}{
        $T_\text{Cali} = T$ 
    }
}
Initialize $Group[k]$ \\
\For{$g \in G$}{
    $k = \left\lfloor{T_{\text{drift}, p_{\text{tar}}}[g] / T_\text{Cali}}\right\rfloor$ \\
    Add $g$ into $Group[k]$
}
\Return $Group[k]$
\end{algorithm}

The optimal \(T_\text{Cali}\) tends to occur when some \(T_{\text{drift}, p_{\text{th}}}[g]\) values align with integer multiples of \(T_\text{Cali}\).
 If alignment does not occur, \(T_\text{Cali}\) can be increased without altering the grouping, thereby reducing the calibration frequency as described in \cref{eq:frequency}. To determine the optimal value, we employ \cref{alg: grouping}, traversing the values of \(T_{\text{drift}, p_\text{tar}}[g] / k\) for each gate, particularly those slightly smaller than the minimum drift time \(T_\text{min}\). The \(T_\text{Cali}\) that minimizes the calibration frequency is then selected.  In cases where multiple intervals yield similar or identical calibration frequencies, a larger interval is preferred. A larger \(T_\text{Cali}\) allows for grouping more gates together, providing longer scheduling windows, increased flexibility for intra-group scheduling, and greater opportunities for parallelism.

\noindent\textbf{Targeted Physical Error Rate Determination:}  \cref{alg: grouping} assumed that the targeted physical error rate, \(p_{\text{tar}}\), was known. Here we describe how our compiler determines \(p_{\text{tar}}\) based on the available physical qubit resources. Surface codes with larger code distances possess stronger error correction capabilities, allowing them to tolerate higher \(p_{\text{tar}}\) while maintaining the same targeted logical error rate (\(\text{LER}_{\text{tar}}\)). 

The LER for a distance-\(d\) surface code is given by~\cite{fowler2018low}:  
\begin{equation}
    \label{eq: LER}
    \text{LER}(d, p_{\text{tar}}) = \alpha \left(\frac{p_{\text{tar}}}{p_{\text{th}}}\right)^{(d+1)/2},
\end{equation}
where \(\alpha\) is a constant specific to the quantum error correction (QEC) code, typically around 0.03 for the rotated surface code. Here, \(p_{\text{th}}\) represents the physical error threshold for the surface code, approximately 0.01 under the circuit-level noise model.

To guarantee program fidelity, the condition \(\text{LER} \leq \text{LER}_{\text{tar}}\) must be satisfied. With~\cref{eq: Drift Constant Time}~and~\cref{eq: LER}, this condition can be simplified:  
\begin{equation}
    \label{eq: constraint}
    \left(\log \frac{p_{\text{th}}}{p_0} - \log \frac{p_{\text{tar}}}{p_0}\right) \cdot d \geq \lambda,
\end{equation}
where \(\lambda\) is a positive constant dependent on \(\text{LER}_{\text{tar}}\) and \(\alpha\). Importantly, this condition can only be satisfied if \(p_{\text{tar}} < p_{\text{th}}\), which aligns with the intuitive requirement that no gate’s error rate should exceed the surface code’s physical error threshold during program execution.  

Given a fixed number of physical qubits, the compiler calculates the maximum allowable code distance \(d\) that fits within the resource constraints. It then determines the largest \(p_{\text{tar}}\) that satisfies the target \(\text{LER}_{\text{tar}}\) while ensuring \(\text{LER}(t) \leq \text{LER}_{\text{tar}}\) throughout program execution.  

This approach balances the trade-offs between code distance, \(p_{\text{tar}}\), and physical qubit resources. Larger code distances exponentially suppress logical error rates but require more physical qubits. Conversely, higher \(p_{\text{tar}}\) allows for longer drift times and less frequent calibration but relies on more robust error correction. By leveraging these trade-offs, our compiler optimizes \(p_{\text{tar}}\) and \(d\) to achieve reliable and resource-efficient execution of quantum programs.
\subsection{Intra-Group Calibration Scheduling:}
\label{subsec:technical-subsec3}

After assigning calibrations to each time period \(T_{\text{Cali}}\) and determining which gates should be calibrated simultaneously, the next step involves finding an appropriate code deformation process to isolate these gates. Generally, for each gate, we apply code deformer to isolate all its affected neighboring qubits, enabling a region suitable for calibration. However, performing this sequentially may lead to excessive calibration time overhead, potentially causing the calibration time of a gate, \(t_{\text{cali}}\), to exceed the calibration cycle \(T_{\text{Cali}}\). 

In this section, we analyze three challenges associated with calibration scheduling and propose an adaptive calibration scheduling approach to address them.

\begin{figure*}[t]
    \centering
    \includegraphics[width=0.90\textwidth]{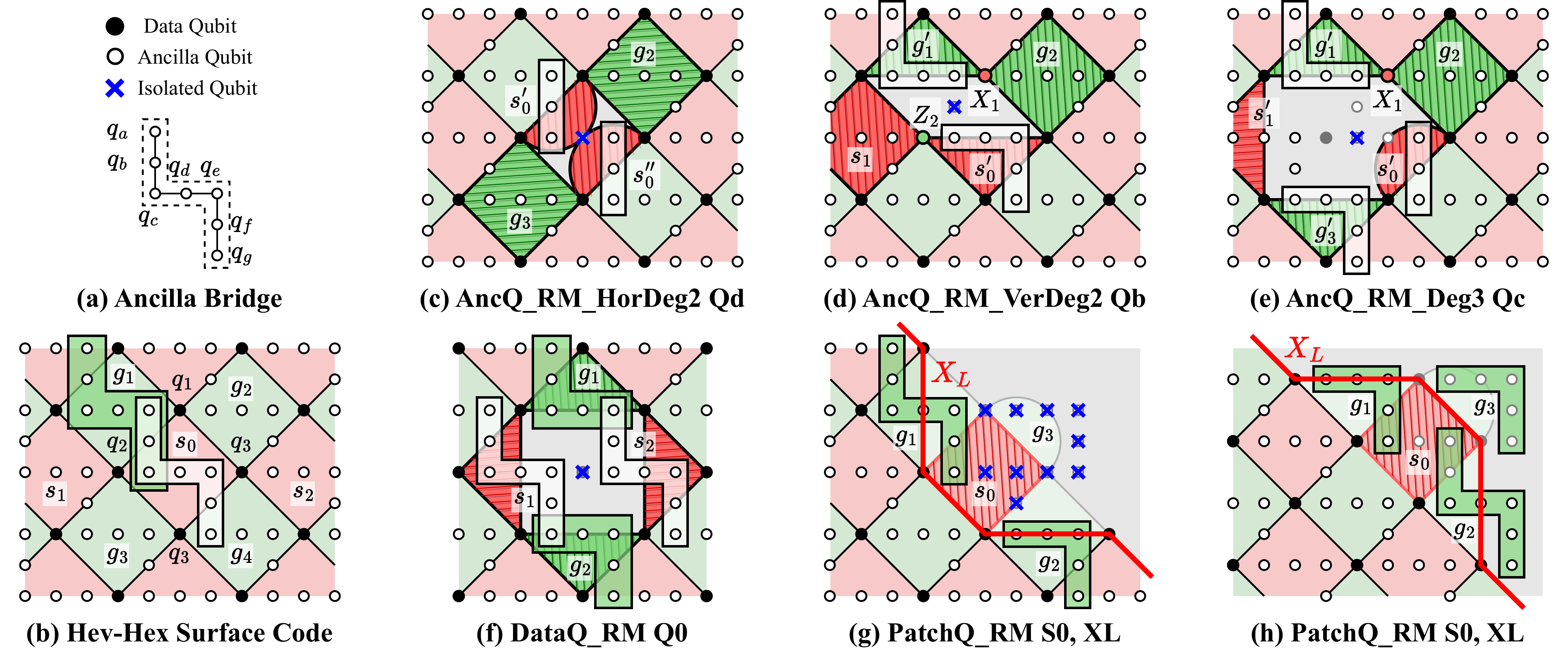}
    \vspace{-8pt}
    \caption{Code deformation instruction set for the surface code on the heavy-hexagon structure.}
    \label{fig: RmSynd}
\end{figure*}

\textbf{(1) Dependence between calibrations:}  
Certain two-qubit gate calibrations may depend on the results of one-qubit gate calibrations. We address this by clustering such gates and scheduling their calibration collectively. This dependency typically arises from overlapping positions, meaning their affected neighboring qubits \(nbr(g)\) are highly overlapped. Code deformation to isolate these shared qubits facilitates the simultaneous calibration of multiple gates in the cluster.

\textbf{(2) Crosstalk between calibrations:}  
Crosstalk during calibrations prevents all calibrations from running simultaneously. To maximize calibration parallelism and minimize calibration time, we design a greedy scheduling (bulk strategy). This approach sorts gate calibrations based on the size of their affected qubits. Starting with the largest, we select as many calibrations as possible without introducing crosstalk. When no more gates can be calibrated concurrently, we generate code deformation instructions for the selected gates and begin a new batch. This process is repeated until all gates are calibrated.

\textbf{(3) Trade-off between calibration time and code distance loss:}  
Code deformation inevitably reduces the code distance by \(\Delta d\). To preserve the fidelity of the original code, the code must be enlarged to restore the original code distance, as discussed in \cite{yin2024surf}. Calibrating more gates simultaneously requires isolating larger regions, leading to greater distance loss and increased physical qubit overhead for enlargement. 

To balance this trade-off, we define a new metric to evaluate scheduling: the space-time overhead of the enlarged region:  

\begin{equation}
    \text{Cost} = \Delta d \sum_g t_{\text{cali}}[g]
\end{equation}
For each \(\Delta d\), we calculate the cost by constraining the greedy scheduling strategy to ensure that the code deformation for simultaneous calibrations does not exceed the allowed distance loss \(\Delta d\). The optimal scheduling is then chosen based on this evaluation.

\section{ The \frameworkName~Instruction Set }
\label{sec: instruction set}

A critical challenge in runtime calibration is the ability to isolate physical qubits for calibration without compromising the surface code's error correction capability. To address this challenge, ~\frameworkName~provides carefully designed instruction sets that enable safe transformation of the code structure during calibration.

\cref{tab: instruction set} provides an overview of the two instruction sets. For surface codes on square lattices, ~\frameworkName~adopts the instruction set from~\cite{yin2024surf}, originally designed for handling defective qubits in surface codes. We identified that these instructions—\verb|DataQ_RM|, \verb|SyndromeQ_RM|, \verb|PatchQ_RM|, and \verb|PatchQ_AD|—can be repurposed for calibration, enabling selective qubit isolation while preserving error correction properties. For the details of these instructions, we refer readers to~\cref{sec deformation} or~\cite{yin2024surf}.

\subsection{The \frameworkName~Instruction Set for the Heavy-Hexagon Topology}
Modern quantum processors, particularly IBM's devices, also employ a heavy-hexagon topology (\cref{fig background}(d)). Existing code deformation instruction sets are tailored to square lattice and are incompatible with heavy-hexagon. To address this gap for calibration, we have designed and formalized a dedicated instruction set specifically for the heavy-hexagon topology. 

\begin{table}[ht]
    \centering
    \caption{\frameworkName~instruction sets for square and heavy-hexagon surface codes}
    \vspace*{-0.1cm}
    \resizebox{0.45\textwidth}{!}{  
        \begin{tabular}{|c|c|}
        \hline
            \rowcolor{lightgray} \rule[-1.5ex]{0pt}{4ex}  Code Topology &  Instructions \\
            
            \hline \rule[-1.1ex]{0pt}{2ex} 

            \multirow{3}{*}{
                \makecell{
                    Square
                }
            } 
            & \multirow{3}{*}{
                \makecell{
                    \texttt{DataQ\_RM},~~\texttt{SyndromeQ\_RM}, \\ 
                    \texttt{PatchQ\_RM},~~\texttt{PatchQ\_AD}
                }
            } \\
            \multirow{3}{*}{} & \multirow{3}{*}{} \\
            \multirow{3}{*}{} & \multirow{3}{*}{} \\

            \hline \rule[-1.ex]{0pt}{2ex} 

            \multirow{3}{*}{
                \makecell{
                    Heavy-Hexagon
                }
            } 
            & \multirow{3}{*}{
                \makecell{
                    \texttt{DataQ\_RM},~~\texttt{AncQ\_RM\_HorDeg2}, \\
                    \texttt{PatchQ\_RM},~~\texttt{AncQ\_RM\_VerDeg2}\\
                    \texttt{PatchQ\_AD},~~\texttt{AncQ\_RM\_Deg3}, \\
                }
            } \\
            \multirow{3}{*}{} & \multirow{3}{*}{} \\
            \multirow{3}{*}{} & \multirow{3}{*}{} \\

        \hline
        \end{tabular}
    }
    \label{tab: instruction set}
   \vspace*{-0.2cm}
\end{table}

\noindent \textbf{Key Distinctions of the Topology:} The heavy-hexagon topology has two key features that render the square lattice instructions inadequate:

\noindent \textit{1. Non-uniform ancilla roles:}  
As shown in \cref{fig: RmSynd}(a), the seven ancilla qubits associated with each stabilizer can be classified into two distinct types: (1) Degree-3 nodes (\(q_a, q_c, q_e, q_g\)), which connect to three other qubits, including one data qubit.  (2) Degree-2 nodes (\(q_d, q_b, q_f\)), which connect to two other ancilla qubits.
Degree-3 nodes directly connect to data qubits, while degree-2 nodes act as bridges to link them. This functional difference necessitates different code deformation instructions to ensure the deformed code remains functional after isolating specific ancilla qubits.

\noindent \textit{2. Shared ancilla qubits:}  
As shown in \cref{fig: RmSynd}(a), ancilla qubits associated with one stabilizer can also be shared with others. For instance, in \cref{fig: RmSynd}(b), the \(X\)-stabilizer \(s_0\) and the \(Z\)-stabilizer \(g_1\) share three ancilla qubits. This overlap implies that isolating a single ancilla qubit may impact multiple stabilizers simultaneously which makes designing code deformation instructions under these circumstances becomes more complex.

\noindent \textbf{Design Principles:} The heavy-hexagon structure presents unique opportunities for code deformation compared to the square lattice. We outline the key principle that guided the design of our instruction set --- \noindent \textit{leveraging residual connectivity:} when an ancilla qubit is removed, the remaining structure often retains partial connectivity between data qubits. By utilizing this residual connectivity, we replace the original stabilizer with a product of smaller, localized measurements. This approach minimizes disruption to the overall code structure and preserves a greater number of stabilizers, ensuring robust error correction.

\noindent \textbf{The Instructions:}  
Building on the design principles outlined above, we redesign three square lattice instructions to adapt them to the heavy-hexagon architecture. Additionally, we introduce three distinct instructions of \texttt{AncQ\_RM} category, each specifically tailored to remove ancilla qubits based on their unique positions and connectivity within the heavy-hexagon device. 

\noindent \textbf{(1) \texttt{AncQ\_RM\_HorDeg2}:} This instruction targets a degree-2 horizontal ancilla qubit \(q_d\), as illustrated in \cref{fig: RmSynd}(c). Removing \(q_d\) divides the \(X\)-stabilizer \(s_0\) into two parts, \(s_0' = X_{1,2}\) and \(s_0'' = X_{3,4}\), forming two new gauge measurements that replace the original stabilizer \(s = s_0' s_0''\). Additionally, the removal transforms the nearby \(Z\)-stabilizers \(g_2\) and \(g_3\) into new gauges, which combine to form a new \(Z\)-super-stabilizer \(g_2 g_3\).

\noindent \textbf{(2) \texttt{AncQ\_RM\_VerDeg2}:} This instruction removes a degree-2 vertical ancilla qubit \(q_b\), as shown in \cref{fig: RmSynd}(d). Unlike the horizontal case, \(q_b\) is shared by the \(X\)-stabilizer \(s_0\) and the \(Z\)-stabilizer \(g_1\). Removing \(q_0\) divides \(s_0\) into a three-qubit gauge \(s_0' = X_{2,3,4}\) and a single-qubit gauge \(X_1\). Similarly, \(g_1\) is divided into a three-qubit gauge \(g_1' = Z_{5,6,1}\) and a single-qubit gauge \(Z_2\). These divisions also affect the nearby \(Z\)-stabilizer \(g_2\) and \(X\)-stabilizer \(s_1\), transforming them into new gauges. Collectively, these changes result in a new \(X\)-super-stabilizer \(X_1 s_0' s_1\) and a new \(Z\)-super-stabilizer \(Z_2 g_1' g_2\).

\noindent \textbf{(3) \texttt{AncQ\_RM\_Deg3}:} This instruction removes a degree-3 ancilla qubit \(q_c\), as illustrated in \cref{fig: RmSynd}(e). Similar to \texttt{AncQ\_RM\_VerDeg2}, \(q_e\) is shared by the \(X\)-stabilizer \(s_0\) and the \(Z\)-stabilizer \(g_1\). The removal of \(q_0\) leaves \(g_1\) unchanged, dividing it into two parts: \(g_1' = Z_{5,6,1}\) and a single-qubit gauge \(Z_2\). However, removing \(q_0\) divides \(s_0\) into three components: a two-qubit gauge \(s_0' = X_{3,4}\) and two single-qubit gauges \(X_1\) and \(X_2\).
After this division, both \(X_2\) and \(Z_2\) exist as single-qubit gauges for qubit \(q_2\). This indicates that \(q_2\) becomes a gauge qubit isolated from the surface code, which should be removed. The removal of \(q_2\) further impacts the nearby \(Z\)-stabilizers \(g_2\) and \(g_3\), as well as the \(X\)-stabilizer \(s_1\), deforming and transforming them into new gauges. Collectively, these modifications result in a new \(X\)-super-stabilizer \(X_1 s_0' s_1'\) and a new \(Z\)-super-stabilizer \(g_1' g_2 g_3'\).

\noindent \textbf{(4) \texttt{DataQ\_RM}, \texttt{PatachQ\_RM}, \texttt{PatachQ\_ADD}:} These three instructions deform the surface code in a manner similar to those for the square lattice (\cref{fig deform}), ensuring that the stabilizers in the deformed code remain unchanged. However, because the heavy-hexagon surface code features unique ancilla bridges in its stabilizer circuits, it is also necessary to deform the ancilla bridges associated with the affected stabilizers during the deformation process.

The new instruction set not only make the deformation compatible with heavy-hexagon structures but also leverage the structure of the ancilla bridge to create more fine-grained strategies.

\section{Experimental Setup}
\label{subsec: setup}
\subsection{Setting and benchmark}
\noindent\textbf{Evaluation setting.}
We evaluate QECali through both hardware experiments and simulation-based analysis. Our hardware experiments are conducted on two quantum processors with distinct topologies: Rigetti's Ankaa-2 processor with square lattice connectivity and IBM's Eagle processor with heavy-hex connectivity. For large-scale logical error analysis, we employ the Stim quantum error simulator for surface code simulation~\cite{stim}~along with Pymatching~\cite{higgott2022pymatching}~for error correction. Program compilation utilizes the lattice surgery framework~\cite{herr2017lattice}~and implements logical T gates through magic state distillation~\cite{fowler2018low}.


\vspace{3pt}
\noindent \textbf{Benchmark programs.} We evaluate QECali using quantum programs designed for most promising applications in quantum chemistry and materials science. Our benchmarks include Hubbard model simulation~\cite{arovas2022hubbard}, which provides essential insights into strongly correlated electronic systems with direct applications to high-temperature superconductivity; FeMo-co catalyst analysis~\cite{femoco}~, which addresses the critical industrial challenge in nitrogen fixation; and Jellium simulation~\cite{jellium}~, which serves as a fundamental model for understanding electronic structure in materials. Program variants are denoted by suffixes indicating problem size (e.g., Hubbard-16 for a 16-qubit system).


\vspace{3pt}
\noindent \textbf{Metric.} 
We evaluate QECali using four metrics. The \emph{physical qubit count} encompasses all qubits required for the quantum program, including data qubit blocks for logical encoding, ancilla qubits for CX operations, and resource states for T gate implementation. \emph{Execution time} measures the total runtime for program completion, with QEC cycle time set to $1 \mu s$ (standard in FTQC studies [10, 35, 52, 56]), including all quantum operations and error correction cycles. \emph{Retry risk}~\cite{Gidney2021howtofactorbit}~quantifies the probability of encountering uncorrectable logical errors, providing a measure of program reliability and the likelihood of requiring computation restart.
\emph{Logical error rate (LER)} represents the probability of logical errors occurring per quantum error correction (QEC) cycle for a logical surface code qubit. It reflects the effectiveness of error correction for physical errors, with a lower LER indicating superior fault-tolerant performance.

\begin{figure}[!ht]
    \centering
    \includegraphics[width=0.47\textwidth]{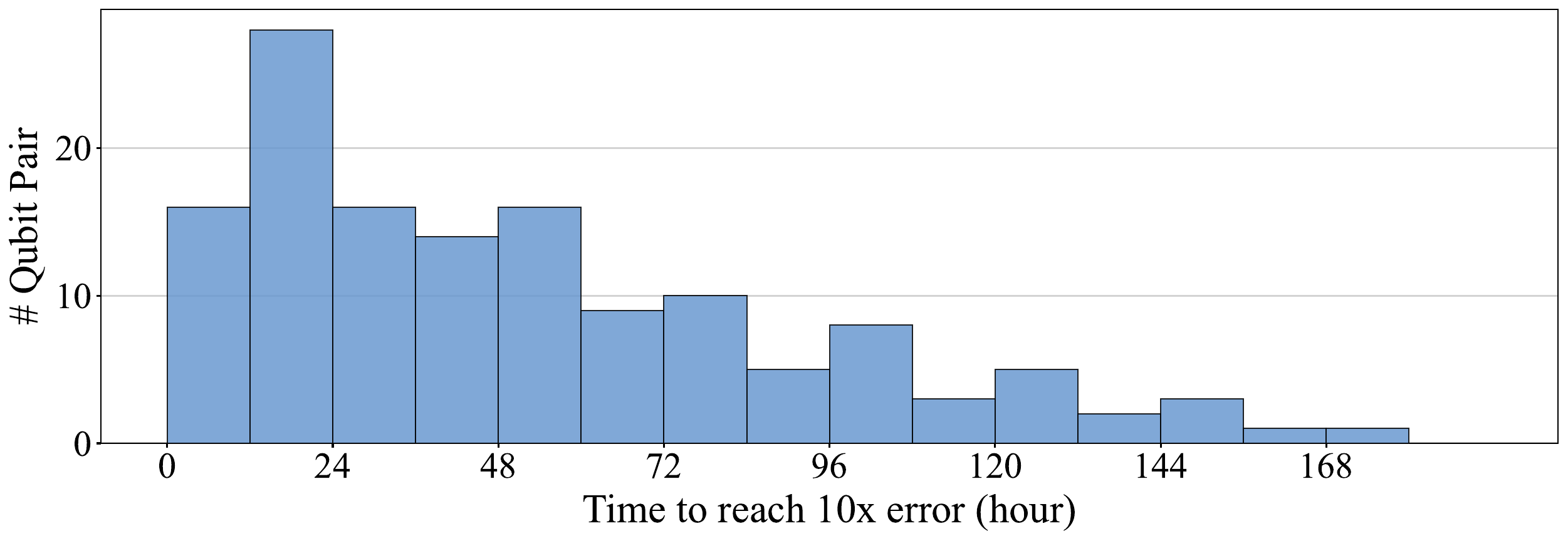}
    \caption{Probability distribution of error drift.}
    \label{fig:drift distribution}
\end{figure}

\subsection{Error model} {\label{subsec: error model}}
\noindent \textbf{Physical error model.}
We adopt a standard circuit-level noise model ~\cite{fowler2012surface, stim} where quantum operations are subject to different error channels: single-qubit gates experience depolarizing errors, two-qubit gates undergo two-qubit depolarizing errors, and measurement and reset operations are affected by Pauli-X errors, each with probability $p$. Initially, all operations start with a uniform error rate \( p = 10^{-X} \), where X is chosen to be 10× below the surface code threshold (1\%). This initialization represents an ideally calibrated device state.


\vspace{3pt}
\noindent \textbf{Error Drift Model.}
Based on our measurements of IBM's 127-qubit Eagle processor, we observe that physical error rates increase exponentially over time, following the relation:
\begin{equation*}
    p(G, t) = p(G, 0) 10^{t / T(G)}
\end{equation*}
where \( p(G, t) \) is the error rate of operation or qubit \(G \) at time \( t \), \( p(G, 0) \) is the initial calibrated error rate, and \( T(G) \) is the  operation-specific drift time constant---the time required for the error rate to increase by a factor of 10.  Our characterization shows that these drift time constants vary significantly across the device, following a log-normal distribution with a mean of 14.08 hours~(\cref{fig:drift distribution}). This heterogeneity stems from variations in qubit connectivity, gate implementation complexity, and device characteristics, necessitating individual calibration schedules for different device components.

\noindent\textbf{Future Error Model. }We account for potential advancements in hardware technology that may slow error drift or, equivalently, extend the constant $T(G)$. To model this, we assume a future error scenario where $T(G)$ follows a log-normal distribution with a doubled mean of 28.016. Our framework is evaluated under this future error model to demonstrate its adaptability and continued utility in evolving hardware environments.

\label{sec:evaluation}

\def \ncol{3}
\def \nhead{5}
\def \endcol{15}
\begin{table*}[!htbp]
    \centering
    \caption{Comparison of performance for large scale programs}
    
    \resizebox{0.98\textwidth}{!}{
        \begin{tabular}{
            |c
            |*{\nhead}{|c} | 
            *{\ncol}{|c} | 
            *{\ncol}{|c} | 
            *{\ncol}{|c} ||
        } 
        
        \hline
        \multirow{3}{*}{\makecell{Error drift\\Model}}
        &\multicolumn{\nhead}{c||}{Benchmark}
        & \multicolumn{\ncol}{c||}{No-Change} 
        & \multicolumn{\ncol}{c||}{LSC}
        & \multicolumn{\ncol}{c||}{\frameworkName} 
        \\
        
        \cline{2-\endcol}
        &\multirow{2}{*}{Name} 
        & \multirow{2}{*}{\# $CX$}
        & \multirow{2}{*}{\# $T$}
        & \multirow{2}{*}{\makecell{\# logical \\ qubit}}
        & \multirow{2}{*}{d}
        
        
        & \multirow{2}{*}{\makecell{\# physical \\ qubit}}
        & \multirow{2}{*}{\makecell{ Execution \\ time}}
        & \multirow{2}{*}{\makecell{Retry \\ risk}}

        & \multirow{2}{*}{\makecell{\# physical \\ qubit}}
        & \multirow{2}{*}{\makecell{ Execution \\ time}}
        & \multirow{2}{*}{\makecell{Retry \\ risk}}

        & \multirow{2}{*}{\makecell{\# physical \\ qubit}}
        & \multirow{2}{*}{\makecell{ Execution \\ time}}
        & \multirow{2}{*}{\makecell{Retry \\risk}}
        \\
        
        & & & & &
        & & &  
        & & &  
        & & &  
        \\
        
        \hline
        \multirow{10}{*}{\makecell{Current \\ model}}
        & \multirow{2}{*}{\makecell{Hubbard\\-10-10}} 
        & \multirow{2}{*}{$1.64 \times 10^9$} 
        & \multirow{2}{*}{$7.10 \times 10^8$}
        & \multirow{2}{*}{$200$} 
        & 25 
        & $9.81 \times 10^5$ & $5.29$ &$\sim 100\%$
        & $4.65 \times 10^6$ & $5.74$ & $11.3\%$
        & $1.53 \times 10^6$ & $5.29$ & $3.13\%$
        \\
        
        \cline{6-\endcol} 
        &  &  &  &
        & 27
        & $1.14 \times 10^6$ & $5.50$ &$\sim 100\%$
        & $5.43 \times 10^6$ & $5.95$ & $1.22\%$
        & $1.62 \times 10^6$ & $5.50$ & $0.38\%$
        \\
        
        \cline{2-\endcol}

        & \multirow{2}{*}{\makecell{Hubbard\\-20-20}} 
        & \multirow{2}{*}{$5.3 \times 10^{10}$} 
        & \multirow{2}{*}{$1.2 \times 10^{10}$}
        & \multirow{2}{*}{$800$} 
        & 29
        & $5.28 \times 10^6$ & $91.3$ &$\sim 100\%$
        & $2.30 \times 10^7$ & $101.5$ & $7.35\%$
        & $7.11 \times 10^6$ & $91.3$ & $1.88\%$
        \\
        
        \cline{6-\endcol} 
        &  &  &  &
        & 31
        & $6.03 \times 10^6$ & $94.3$ &$\sim 100\%$
        & $2.63 \times 10^7$ & $108.5$ & $0.79\%$
        & $8.38 \times 10^6$ & $94.3$ &  $0.20\%$
        \\
        
        \cline{2-\endcol}

        &\multirow{2}{*}{\makecell{jellium\\-250}} 
        & \multirow{2}{*}{$8.23 \times 10^9$} 
        & \multirow{2}{*}{$1.10 \times 10^9$}
        & \multirow{2}{*}{$250$} 
        & 39 
        & $2.74 \times 10^6$ & $177$ &$\sim 100\%$
        & $1.29 \times 10^7$ & $190.5$ & $8.65\%$
        & $4.87 \times 10^6$ & $177$ & $2.40\%$
        \\
        
        \cline{6-\endcol} 
        & &  &  &
        & 41
        & $3.03 \times 10^6$ & $182$ &$\sim 100\%$
        & $1.42 \times 10^7$ & $195.95$ & $0.91\%$
        & $5.38 \times 10^6$ & $182$ & $0.24\%$
        \\
        
        \cline{2-\endcol}

        &\multirow{2}{*}{\makecell{jellium\\-1024}} 
        & \multirow{2}{*}{$1.25 \times 10^{12}$} 
        & \multirow{2}{*}{$4.30 \times 10^{10}$}
        & \multirow{2}{*}{$1024$} 
        & 45
        & $1.66 \times 10^7$ & $1870$ &$\sim 100\%$
        & $7.17 \times 10^7$ & $2010.4$ & $3.69\%$
        & $2.22 \times 10^7$ & $1870$ & $0.88\%$
        \\
        
        \cline{6-\endcol} 
        & &  &  &
        & 47
        & $1.81 \times 10^7$ & $2140$ &$\sim 100\%$
        & $7.82 \times 10^7$ & $2300$ & $0.39\%$
        & $2.42 \times 10^7$ & $2140$ & $0.09\%$
        \\
        
        \cline{2-\endcol}

        &\multirow{2}{*}{Grover-100} 
        & \multirow{2}{*}{$6.8 \times 10^9$} 
        & \multirow{2}{*}{$5.4 \times 10^{10}$}
        & \multirow{2}{*}{$100$} 
        & 41
        & $1.35 \times 10^6$ & $220$ &$\sim 100\%$
        & $6.81 \times 10^6$ & $236.5$ & $6.16\%$
        & $3.03 \times 10^6$ & $220$ & $0.98\%$
        \\
        
        \cline{6-\endcol} 
        & &  &  &
        & 43
        & $1.48 \times 10^6$ & $237$ &$\sim 100\%$
        & $7.49\times 10^6$ & $243.67$ & $0.92\%$
        & $3.33 \times 10^6$ & $237$ & $0.11\%$
        \\
        
        \hline
        \multirow{10}{*}{\makecell{Future\\ model}}
        & \multirow{2}{*}{\makecell{Hubbard\\-10-10}} 
        & \multirow{2}{*}{$1.64 \times 10^9$} 
        & \multirow{2}{*}{$7.10 \times 10^8$}
        & \multirow{2}{*}{$200$} 
        & 25 
        & $9.81 \times 10^5$ & $5.29$ &$\sim 100\%$
        & $4.65 \times 10^6$ & $5.29$ & $3.13\%$
        & $1.36 \times 10^6$ & $5.29$ & $3.13\%$
        \\
        
        \cline{6-\endcol} 
        & &  &  &
        & 27
        & $1.14 \times 10^6$ & $5.50$ &$\sim 100\%$
        & $5.43 \times 10^6$ & $5.50$ & $0.38\%$
        & $1.59 \times 10^6$ & $5.50$ & $0.38\%$
        \\
        
        \cline{2-\endcol}

        &\multirow{2}{*}{\makecell{Hubbard\\-20-20}} 
        & \multirow{2}{*}{$5.3 \times 10^{10}$} 
        & \multirow{2}{*}{$1.2 \times 10^{10}$}
        & \multirow{2}{*}{$800$} 
        & 29
        & $5.28 \times 10^6$ & $91.3$ &$\sim 100\%$
        & $2.30 \times 10^7$ & $94.7$ & $7.35\%$
        & $6.26 \times 10^6$ & $91.3$ & $1.88\%$
        \\
        
        \cline{6-\endcol} 
        & &  &  &
        & 31
        & $6.03 \times 10^6$ & $94.3$ &$\sim 100\%$
        & $2.63 \times 10^7$ & $97.8$ & $0.79\%$
        & $7.16 \times 10^6$ & $94.3$ &  $0.20\%$
        \\
        
        \cline{2-\endcol}

        &\multirow{2}{*}{\makecell{jellium\\-250}} 
        & \multirow{2}{*}{$8.23 \times 10^9$} 
        & \multirow{2}{*}{$1.10 \times 10^9$}
        & \multirow{2}{*}{$250$} 
        & 39 
        & $2.74 \times 10^6$ & $177$ &$\sim 100\%$
        & $1.29 \times 10^7$ & $183.3$ & $8.65\%$
        & $3.73 \times 10^6$ & $177$ & $2.40\%$
        \\
        
        \cline{6-\endcol} 
        & &  &  &
        & 41
        & $3.03 \times 10^6$ & $182$ &$\sim 100\%$
        & $1.42 \times 10^7$ & $188.83$ & $0.91\%$
        & $4.12 \times 10^6$ & $182$ & $0.24\%$
        \\
        
        \cline{2-\endcol}

        &\multirow{2}{*}{\makecell{jellium\\-1024}} 
        & \multirow{2}{*}{$1.25 \times 10^{12}$} 
        & \multirow{2}{*}{$4.30 \times 10^{10}$}
        & \multirow{2}{*}{$1024$} 
        & 45
        & $1.66 \times 10^7$ & $1870$ &$\sim 100\%$
        & $7.17 \times 10^7$ & $1960$ & $3.69\%$
        & $1.93 \times 10^7$ & $1870$ & $0.88\%$
        \\
        
        \cline{6-\endcol} 
        & &  &  &
        & 47
        & $1.81 \times 10^7$ & $2140$ &$\sim 100\%$
        & $7.82 \times 10^7$ & $2220$ & $0.39\%$
        & $2.10 \times 10^7$ & $2140$ & $0.09\%$
        \\
        
        \cline{2-\endcol}

        & \multirow{2}{*}{Grover-100} 
        & \multirow{2}{*}{$6.8 \times 10^9$} 
        & \multirow{2}{*}{$5.4 \times 10^{10}$}
        & \multirow{2}{*}{100} 
        & 41 
        & $1.35 \times 10^6$ & $220$ &$\sim 100\%$
        & $6.81 \times 10^6$ & $228.25$ & $6.16\%$
        & $2.10 \times 10^6$ & $220$ & $0.98\%$
        \\
        
        \cline{6-\endcol} 
        & &  &  &
        & 43
        & $1.48 \times 10^6$ & $237$ &$\sim 100\%$
        & $7.49 \times 10^6$ & $245.89$ & $0.92\%$
        & $2.31 \times 10^6$ & $237$ & $0.11\%$
        \\
        
        \hline
        
        \end{tabular}
    }
    \label{tab:evaluation}
\end{table*}

\section{Evaluation}
In this section, we evaluate~\frameworkName~by comparing with
two baselines, analyze the effects of individual
components, and perform experiments on real systems.
\subsection{Overall Performance}
We evaluated~\frameworkName~against two baseline approaches: running without calibration and using~\baselineLong~(\baseline), under both the current and future error models described in Sec.~\ref{subsec: error model}. Our experiments, presented in~\cref{tab:evaluation}, use surface codes with distances chosen to achieve target retry risk levels of 1\% and 0.1\% and benchmarks described in~\cref{subsec: setup}. Our evaluation reveals four critical observations:

\noindent\textbf{1. Calibration is indispensable}: attempting to run quantum programs without calibration leads to retry risks approaching 100\%, demonstrating the severe impact of error drift. While programs start with low logical error rates, exponential drift in physical error rates quickly compromises computation reliability, making successful execution virtually impossible for long-running applications.

\noindent\textbf{2. Coarse-grained calibration is impractical}: Compared to the solution with no calibration (the “No-Change” column in Table~\ref{tab:evaluation}), the~\baseline~approach reduces the retry risk to the target level but incurs a substantial 363\% increase in qubit count and a 20\% longer execution time. These inefficiencies stem from~\baseline's need to double qubit count for logical state storage during migration, combined with execution delays from logical SWAP operations and program stalls during calibration.~\baseline's coarse-grained approach, where calibration timing is determined by the worst-performing qubits, further compounds these inefficiencies as system size grows.



\noindent\textbf{3.~\frameworkName~achieves low overhead calibration}: Compared to the~\baseline~solution, \frameworkName~sustains computation progress during calibration, dramatically reducing the 363\% qubit overhead of~\baseline~to just 24\% and eliminating execution time overhead entirely. By performing in-situ calibration via code deformation, \frameworkName~avoids the computational stalls and ancilla overhead associated with state transfer approaches. Furthermore, \frameworkName~reduces retry risk by 79.4\% compared to~\baseline, demonstrating the effectiveness of its fine-grained calibration strategy in preserving the system’s error correction capabilities and suppressing the retry risk.


\noindent\textbf{4.~\textit{In situ} calibration remains essential even with improved hardware}: The lower half of Table~\ref{tab:evaluation} compares \frameworkName~with two baselines under the future error model. In this scenario, \baseline~still incurs a significant qubit count overhead (363\%) because the physical error rate eventually exceeds the threshold, necessitating logical state transfers for calibration and requiring excessive additional qubit resource. While slower error drift reduces the frequency of calibration, \baseline~still suffers from a 304\% higher retry risk, despite a modest increase in execution time. This analysis highlights that even with reduced error drift in future systems, large-scale quantum tasks will still require \textit{in situ} calibration. The log-normal distribution of drift time constants $T(G)$ across gates ensures that some physical qubits will remain more vulnerable to drift. Our experiments confirm that even a small number of underperforming qubits can significantly increase logical error rates, reinforcing the importance of runtime calibration for reliable quantum computation.

\subsection{Component-wise Analysis}
We conducted a detailed analysis of~\frameworkName's key components to quantify their individual contributions to overall system performance. This ablation study focuses particularly on our adaptive calibration scheduling and resource management strategies.


\begin{figure}[!ht]
    \centering
    \includegraphics[width=0.47\textwidth]{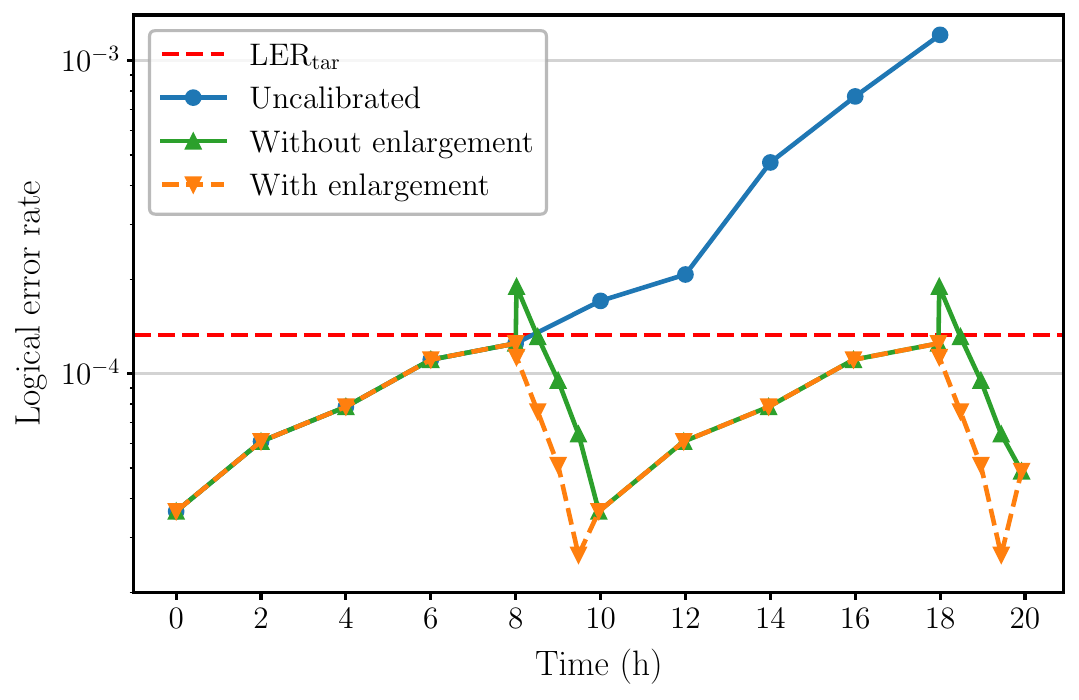}
    \caption{d=11 Logical error rate analysis with error drift.}
    \label{fig: analysis}
\end{figure}

\noindent\textbf{9.2.1\ \ \ In Situ Calibration's Impact on LER}

We evaluate the impact of the two critical deformation steps in our calibration framework: qubit isolation and code enlargement.
In Fig.~\ref{fig: analysis}, we simulate the LER dynamics during calibration cycles for a $d = 11$ surface code. The red line represents the LER threshold, indicating the maximum allowable LER to maintain the desired retry risk level. The blue, green, and orange lines illustrate LER dynamics under three scenarios: (1) no calibration, (2) qubit isolation + calibration, and (3) qubit isolation + code enlargement + calibration. The results demonstrate: (1) Without calibration, the LER increases exponentially due to error drift. (2) With qubit isolation and calibration but no code enlargement, the LER briefly spikes above the threshold due to distance loss from qubit isolation, though it eventually falls below the threshold after calibration. (3) The complete \frameworkName~scheme, incorporating both qubit isolation and code enlargement, quickly restores error protection and keeps the LER consistently below the threshold.

Importantly, this compensation mechanism of~\frameworkName~proves highly efficient: the code distance reduction (\(\Delta d\)) during calibration requires only a \(d + \Delta d\) expansion, resulting in 14\% additional physical qubits. This modest overhead can be further optimized by adjusting calibration intervals to minimize distance loss. Moreover, since compensation qubits are only needed during calibration, they can be shared across different logical qubits through our flexible layout scheme. This sharing reduces the net qubit overhead to 6\%, while maintaining sub-threshold logical error rates throughout computation.


\noindent\textbf{9.2.2\ \ \ Impact of Adaptive Calibration Assignment}

\begin{figure}[!ht]
    \centering
    \includegraphics[width=0.45\textwidth]{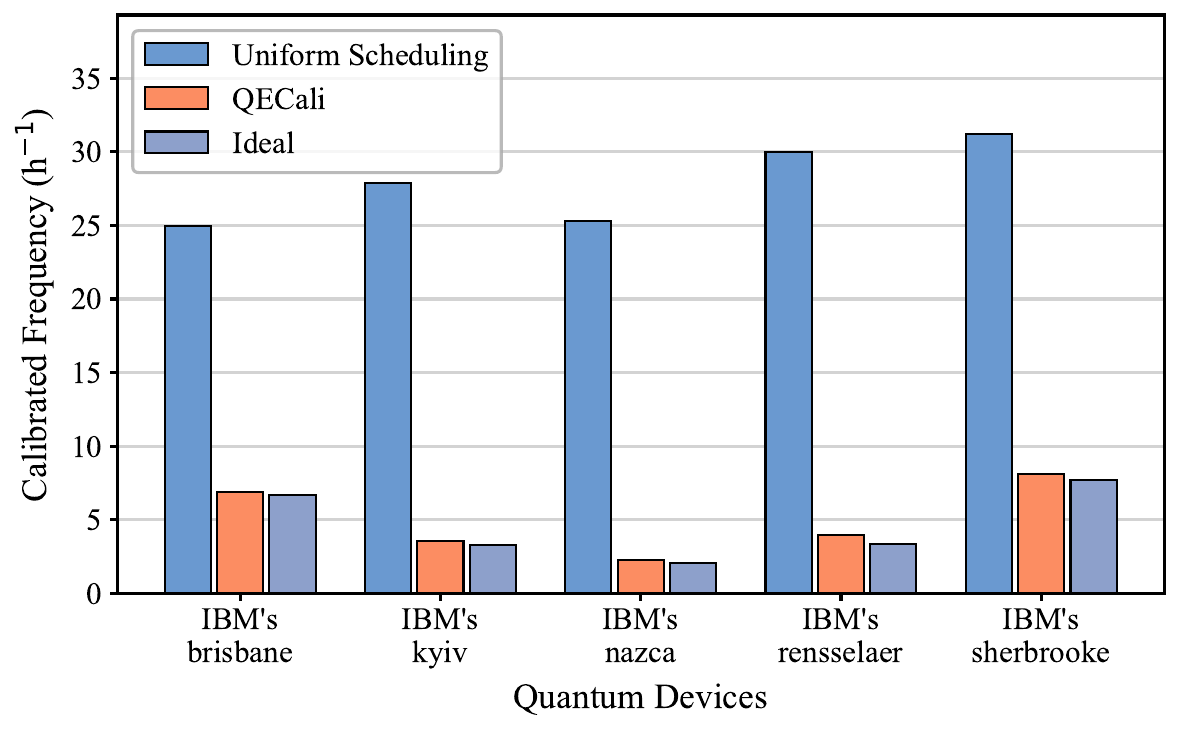}
    \caption{Reduction in calibration count through adaptive calibration assignment}
    \label{fig: Global Scheduling}
\end{figure}

 Naive approaches apply uniform calibration schedules across all qubits, leading to unnecessary calibrations of stable components. In contrast, ~\frameworkName's adaptive approach leverages the natural variation in qubit stability, demonstrated by the normal distribution of drift rates shown in~\cref{fig: Global Scheduling}. By assigning calibration schedules based on individual drift patterns, ~\frameworkName~reduces the total number of calibration operations by 3.63 times to 11.1 times compared to uniform scheduling (\cref{fig: Global Scheduling}). This reduction directly translates to lower operational overhead without compromising error protection.

\noindent\textbf{9.2.3\ \ \ Impact of Adaptive Calibration Scheduling} 

Calibration scheduling must balance speed against qubit overhead: more parallel calibrations reduce execution time but require more compensation qubits. Neither purely sequential nor fully parallel approaches are optimal for practical systems.

To quantify this trade-off, we evaluate scheduling strategies using a space-time overhead metric:
\begin{equation*}
    \text{Overhead} = \Delta d \times T(\text{Cali})
\end{equation*}

where \(\Delta d\) represents the temporary reduction in code distance during calibration, and $T(\text{Cal})$ is the total calibration time. This metric captures both the spatial cost (additional physical qubits needed for code compensation, which scales as O(\(\Delta d\))) and temporal cost (duration of reduced error protection) of the calibration process.

We compare three scheduling approaches: sequential calibration, which processes one gate at a time; maximum parallelism, which calibrates as many gates as dependencies allow; and QECali's adaptive scheduling, which optimizes the parallelism-overhead trade-off. 

\begin{figure}[!ht]
    \centering
    \includegraphics[width=0.45\textwidth]{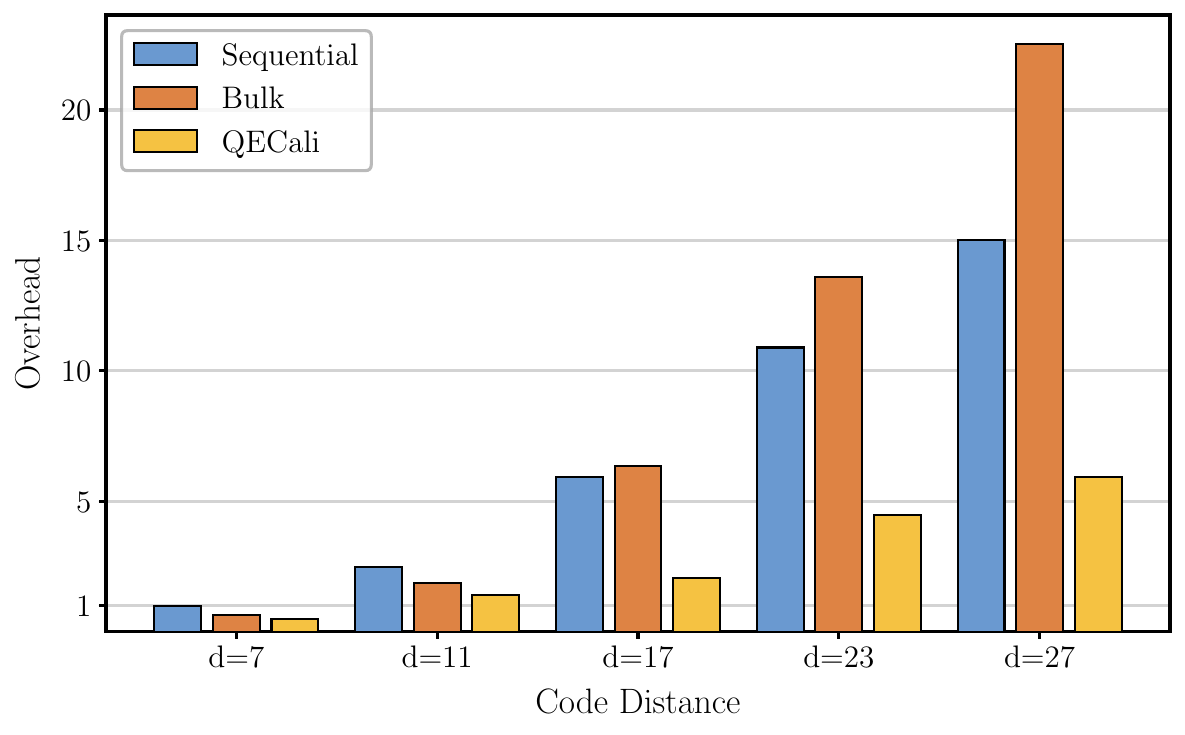}
    \caption{Space-time overhead of calibration of code with different code distance}
        \vspace*{-0.2cm}
    \label{fig: ST overhead}

\end{figure}

Our results (\cref{fig: ST overhead}) show that~\frameworkName~reduces space-time overhead by 2.89 times compared to sequential calibration and 3.8 times compared to maximum parallelism. This improvement demonstrates that naive approaches to parallelization can be counterproductive—either consuming excessive qubit resources (maximum parallelism) or requiring unnecessarily long calibration times (sequential). QECali's adaptive scheduling finds an effective balance, minimizing both resource requirements and calibration duration.

\vspace*{-0.2cm}


\subsection{\frameworkName~on real quantum device}

Real quantum devices present additional challenges: non-uniform gate fidelities, complex error correlations, and hardware-specific constraints. To validate \frameworkName's practicality, we implement $d =3$ surface codes on two state-of-the-art quantum processors with distinct architectures: (1) Rigetti Ankaa-2 with a square lattice architecture, and (2) IBM-Rensselaer with heavy-hexagon architecture. We compare three scenarios: (1) \emph{optimal} (``Original'' column), (2) \emph{drifted}, where a single gate's (either single-qubit or two-qubit gate) calibration parameters are replaced by those drifted after 8 hours (``drifted 1Q'' and ``drifted 2Q'' columns), and (3) \emph{drifted + qubit isolation} (two ``isolated drifted'' columns).
\begin{figure}[!ht]
    \centering
    \includegraphics[width=0.45\textwidth]{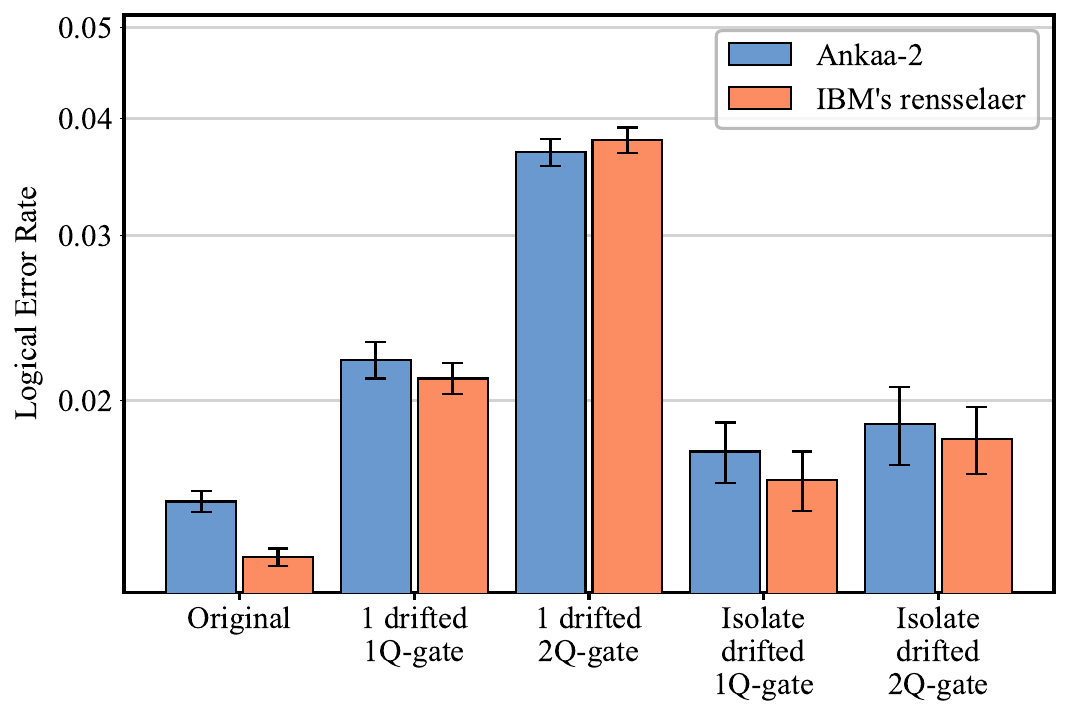}
    \caption{Logical error rate of a $d=3$ surface code on Rigetti Ankaa-2 and IBM-Rensselaer}
    \label{fig: LER under calibration}
\end{figure}

\textbf{Standard surface code on Rigetti Ankaa-2:} 
 Our results (\cref{fig: LER under calibration}) reveal that even a single uncalibrated gate increases the LER by 41.6\% for a single-qubit gate and 135.5\% for a two-qubit gate. In contrast,~\frameworkName’s qubit isolation and recalibration strategy incurs only a minor LER increase during calibration: 13.1\% for single-qubit gates and 21.0\% for two-qubit gates. This demonstrates that isolating high-error-rate qubits affected by error drift effectively suppresses the LER, requiring only a modest qubit overhead for code enlargement to restore QEC capability to the original LER level. Without isolating drifted qubits, significantly more qubits would be needed for enlargement, as the LER difference between drifted qubits (``drifted 1Q'' and ``drifted 2Q'' columns) and the optimal ones (``Original’’ column) is much larger than that between the drift-removed qubits (``isolate drifted 1Q'' and ``isolated drifted 2Q'' columns) and the optimal ones.

\textbf{Heavy-hex surface code on IBM-Rensselaer:} 
We observe similar results on the IBM-Rensselaer device. The drift of a single gate increases the LER by 55.0\% for single-qubit gates and 178.2\% for two-qubit gates. In contrast, removing the drifted qubits limits the LER increase to just 22.8\% and 33.6\%, significantly reducing the qubit resources required to restore the original QEC protection level. Notably, this device is more sensitive to drifted errors than the Rigetti Ankaa-2 device, as evidenced by the larger LER increases (55.0\% and 178.2\% compared to 41.6\% and 135.5\%). We speculate that this heightened sensitivity arises from the heavy-hex topology, where two-qubit gates are shared across multiple stabilizer measurements, amplifying the effect of individual gate errors.

These real-device experiments confirm that qubit isolation is an effective strategy for addressing drifted errors, requiring minimal qubit overhead for code enlargement. By combining qubit isolation with code enlargement, the LER can be kept sufficiently low to safeguard ongoing computation while enabling efficient calibration.


\section{Related Work}
Prior work addressing error drift in quantum systems falls into two main categories: calibration methods and calibration-free methods.

\vspace{3pt}
\noindent\textbf{Calibration Methods. }Calibration methods can be broadly categorized into two types. (1) \emph{Physical methods.} They involve experiments on qubits to determine optimal control parameters. Early methods like tomography~\cite{chuang1997prescription} reconstructed the full quantum state or process matrix through extensive measurements but were impractical for large systems due to exponential scaling. Scalable alternatives such as randomized benchmarking (RB)\cite{knill2008randomized, magesan2011scalable, magesan2012efficient} simplified calibration by measuring average gate fidelity with random gate sequences, becoming a standard technique. Cross-entropy benchmarking (XEB)~\cite{arute2019quantum}, designed for multi-qubit systems, further advanced fidelity assessment by comparing measured outcomes to ideal distributions. Recent approaches leveraging quantum signal processing (QSP)~\cite{martyn2021grand, dong2022beyond} reduce calibration time complexity but require collapsing quantum states, making them incompatible with continuous computation. (2) \emph{Statistical methods.} They estimate parameters using computational data such as historical error syndromes without collapsing quantum states~\cite{kelly2016scalable, wagner2021optimal, wagner2022pauli, wagner2023learning, fowler2014scalable, fujiwara2014instantaneous, huo2017learning}. While they avoid collapsing quantum states or disrupting computation, the derived control parameters are often suboptimal and insufficient for meeting QEC requirements.

\vspace{3pt}
\noindent\textbf{Calibration-free Methods.}  Recent work has also explored adapting error correction strategies to handle drift without explicit calibration. 
 These approaches include dynamic decoder weight adjustment~\cite{etxezarreta2021time, etxezarreta2023multiqubit, wang2023dgr}, which improves error correction temporarily but fails under accelerated drift, temporal noise tracking in decoders to identify drift patterns, and code size adaptation~\cite{suzuki2022q3de, yin2024surf} to reduce logical error rates. However, these methods become ineffective once physical error rates exceed the surface code threshold.

 Unlike these approaches that attempt to work around error drift, QECali directly addresses the underlying physical error rates through in-situ calibration while maintaining computational progress. This is essential for long-term reliability, as all alternative approaches eventually fail when physical error rates exceed the threshold.
\section{Conclusion}

We present~\frameworkName, a framework enabling {\it in-situ} calibration of physical qubits while maintaining quantum error correction in surface codes. Through selective qubit isolation and dynamic code enlargement,~\frameworkName~achieves concurrent calibration and computation while preserving error correction capabilities. Our experimental results demonstrate that~\frameworkName~maintains sub-threshold error rates with only 12-15\% qubit overhead and reduces retry risk by up to 85\% compared to conventional approaches. This effectiveness of code deformation was validated on both square lattice and heavy-hex architectures. As quantum systems scale and computation times increase,~\frameworkName's approach to resource management provides a practical foundation for maintaining reliable quantum error correction during extended computations.

\section*{Acknowledgment}
We thank the anonymous reviewers for their constructive feedback and AWS Cloud Credit for Research. This work is supported in part by NSF 2048144, NSF 2422169, NSF 2427109. This material is based upon work supported by the U.S. Department of Energy, Office of Science, National Quantum Information Science Research Centers, Quantum Science Center (QSC). This research used resources of the Oak Ridge Leadership Computing Facility (OLCF), which is a DOE Office of Science User Facility supported under Contract DE-AC05-00OR22725. This research used resources of the National Energy Research Scientific Computing Center (NERSC), a U.S. Department of Energy Office of Science User Facility located at Lawrence Berkeley National Laboratory, operated under Contract No. DE-AC02-05CH11231. The Pacific Northwest National Laboratory is operated by Battelle for the U.S. Department of Energy under Contract DE-AC05-76RL01830.

\bibliographystyle{ACM-Reference-Format}
\bibliography{TexFile/references}


\begin{thebibliography}{78}


\ifx \showCODEN    \undefined \def \showCODEN     #1{\unskip}     \fi
\ifx \showDOI      \undefined \def \showDOI       #1{#1}\fi
\ifx \showISBNx    \undefined \def \showISBNx     #1{\unskip}     \fi
\ifx \showISBNxiii \undefined \def \showISBNxiii  #1{\unskip}     \fi
\ifx \showISSN     \undefined \def \showISSN      #1{\unskip}     \fi
\ifx \showLCCN     \undefined \def \showLCCN      #1{\unskip}     \fi
\ifx \shownote     \undefined \def \shownote      #1{#1}          \fi
\ifx \showarticletitle \undefined \def \showarticletitle #1{#1}   \fi
\ifx \showURL      \undefined \def \showURL       {\relax}        \fi
\providecommand\bibfield[2]{#2}
\providecommand\bibinfo[2]{#2}
\providecommand\natexlab[1]{#1}
\providecommand\showeprint[2][]{arXiv:#2}

\bibitem[goo(2023)]%
        {google2023suppressing}
 \bibinfo{year}{2023}\natexlab{}.
\newblock \showarticletitle{Suppressing quantum errors by scaling a surface code logical qubit}.
\newblock \bibinfo{journal}{\emph{Nature}} \bibinfo{volume}{614}, \bibinfo{number}{7949} (\bibinfo{year}{2023}), \bibinfo{pages}{676--681}.
\newblock


\bibitem[Acharya et~al\mbox{.}(2024)]%
        {acharya2024quantumerrorcorrectionsurface}
\bibfield{author}{\bibinfo{person}{Rajeev Acharya}, \bibinfo{person}{Laleh Aghababaie-Beni}, \bibinfo{person}{Igor Aleiner}, \bibinfo{person}{Trond~I. Andersen}, \bibinfo{person}{Markus Ansmann}, \bibinfo{person}{Frank Arute}, \bibinfo{person}{Kunal Arya}, \bibinfo{person}{Abraham Asfaw}, \bibinfo{person}{Nikita Astrakhantsev}, \bibinfo{person}{Juan Atalaya}, \bibinfo{person}{Ryan Babbush}, \bibinfo{person}{Dave Bacon}, \bibinfo{person}{Brian Ballard}, \bibinfo{person}{Joseph~C. Bardin}, \bibinfo{person}{Johannes Bausch}, \bibinfo{person}{Andreas Bengtsson}, \bibinfo{person}{Alexander Bilmes}, \bibinfo{person}{Sam Blackwell}, \bibinfo{person}{Sergio Boixo}, \bibinfo{person}{Gina Bortoli}, \bibinfo{person}{Alexandre Bourassa}, \bibinfo{person}{Jenna Bovaird}, \bibinfo{person}{Leon Brill}, \bibinfo{person}{Michael Broughton}, \bibinfo{person}{David~A. Browne}, \bibinfo{person}{Brett Buchea}, \bibinfo{person}{Bob~B. Buckley}, \bibinfo{person}{David~A. Buell}, \bibinfo{person}{Tim Burger}, \bibinfo{person}{Brian
  Burkett}, \bibinfo{person}{Nicholas Bushnell}, \bibinfo{person}{Anthony Cabrera}, \bibinfo{person}{Juan Campero}, \bibinfo{person}{Hung-Shen Chang}, \bibinfo{person}{Yu Chen}, \bibinfo{person}{Zijun Chen}, \bibinfo{person}{Ben Chiaro}, \bibinfo{person}{Desmond Chik}, \bibinfo{person}{Charina Chou}, \bibinfo{person}{Jahan Claes}, \bibinfo{person}{Agnetta~Y. Cleland}, \bibinfo{person}{Josh Cogan}, \bibinfo{person}{Roberto Collins}, \bibinfo{person}{Paul Conner}, \bibinfo{person}{William Courtney}, \bibinfo{person}{Alexander~L. Crook}, \bibinfo{person}{Ben Curtin}, \bibinfo{person}{Sayan Das}, \bibinfo{person}{Alex Davies}, \bibinfo{person}{Laura~De Lorenzo}, \bibinfo{person}{Dripto~M. Debroy}, \bibinfo{person}{Sean Demura}, \bibinfo{person}{Michel Devoret}, \bibinfo{person}{Agustin~Di Paolo}, \bibinfo{person}{Paul Donohoe}, \bibinfo{person}{Ilya Drozdov}, \bibinfo{person}{Andrew Dunsworth}, \bibinfo{person}{Clint Earle}, \bibinfo{person}{Thomas Edlich}, \bibinfo{person}{Alec Eickbusch},
  \bibinfo{person}{Aviv~Moshe Elbag}, \bibinfo{person}{Mahmoud Elzouka}, \bibinfo{person}{Catherine Erickson}, \bibinfo{person}{Lara Faoro}, \bibinfo{person}{Edward Farhi}, \bibinfo{person}{Vinicius~S. Ferreira}, \bibinfo{person}{Leslie~Flores Burgos}, \bibinfo{person}{Ebrahim Forati}, \bibinfo{person}{Austin~G. Fowler}, \bibinfo{person}{Brooks Foxen}, \bibinfo{person}{Suhas Ganjam}, \bibinfo{person}{Gonzalo Garcia}, \bibinfo{person}{Robert Gasca}, \bibinfo{person}{Élie Genois}, \bibinfo{person}{William Giang}, \bibinfo{person}{Craig Gidney}, \bibinfo{person}{Dar Gilboa}, \bibinfo{person}{Raja Gosula}, \bibinfo{person}{Alejandro~Grajales Dau}, \bibinfo{person}{Dietrich Graumann}, \bibinfo{person}{Alex Greene}, \bibinfo{person}{Jonathan~A. Gross}, \bibinfo{person}{Steve Habegger}, \bibinfo{person}{John Hall}, \bibinfo{person}{Michael~C. Hamilton}, \bibinfo{person}{Monica Hansen}, \bibinfo{person}{Matthew~P. Harrigan}, \bibinfo{person}{Sean~D. Harrington}, \bibinfo{person}{Francisco J.~H. Heras},
  \bibinfo{person}{Stephen Heslin}, \bibinfo{person}{Paula Heu}, \bibinfo{person}{Oscar Higgott}, \bibinfo{person}{Gordon Hill}, \bibinfo{person}{Jeremy Hilton}, \bibinfo{person}{George Holland}, \bibinfo{person}{Sabrina Hong}, \bibinfo{person}{Hsin-Yuan Huang}, \bibinfo{person}{Ashley Huff}, \bibinfo{person}{William~J. Huggins}, \bibinfo{person}{Lev~B. Ioffe}, \bibinfo{person}{Sergei~V. Isakov}, \bibinfo{person}{Justin Iveland}, \bibinfo{person}{Evan Jeffrey}, \bibinfo{person}{Zhang Jiang}, \bibinfo{person}{Cody Jones}, \bibinfo{person}{Stephen Jordan}, \bibinfo{person}{Chaitali Joshi}, \bibinfo{person}{Pavol Juhas}, \bibinfo{person}{Dvir Kafri}, \bibinfo{person}{Hui Kang}, \bibinfo{person}{Amir~H. Karamlou}, \bibinfo{person}{Kostyantyn Kechedzhi}, \bibinfo{person}{Julian Kelly}, \bibinfo{person}{Trupti Khaire}, \bibinfo{person}{Tanuj Khattar}, \bibinfo{person}{Mostafa Khezri}, \bibinfo{person}{Seon Kim}, \bibinfo{person}{Paul~V. Klimov}, \bibinfo{person}{Andrey~R. Klots}, \bibinfo{person}{Bryce Kobrin},
  \bibinfo{person}{Pushmeet Kohli}, \bibinfo{person}{Alexander~N. Korotkov}, \bibinfo{person}{Fedor Kostritsa}, \bibinfo{person}{Robin Kothari}, \bibinfo{person}{Borislav Kozlovskii}, \bibinfo{person}{John~Mark Kreikebaum}, \bibinfo{person}{Vladislav~D. Kurilovich}, \bibinfo{person}{Nathan Lacroix}, \bibinfo{person}{David Landhuis}, \bibinfo{person}{Tiano Lange-Dei}, \bibinfo{person}{Brandon~W. Langley}, \bibinfo{person}{Pavel Laptev}, \bibinfo{person}{Kim-Ming Lau}, \bibinfo{person}{Loïck~Le Guevel}, \bibinfo{person}{Justin Ledford}, \bibinfo{person}{Kenny Lee}, \bibinfo{person}{Yuri~D. Lensky}, \bibinfo{person}{Shannon Leon}, \bibinfo{person}{Brian~J. Lester}, \bibinfo{person}{Wing~Yan Li}, \bibinfo{person}{Yin Li}, \bibinfo{person}{Alexander~T. Lill}, \bibinfo{person}{Wayne Liu}, \bibinfo{person}{William~P. Livingston}, \bibinfo{person}{Aditya Locharla}, \bibinfo{person}{Erik Lucero}, \bibinfo{person}{Daniel Lundahl}, \bibinfo{person}{Aaron Lunt}, \bibinfo{person}{Sid Madhuk}, \bibinfo{person}{Fionn~D.
  Malone}, \bibinfo{person}{Ashley Maloney}, \bibinfo{person}{Salvatore Mandrá}, \bibinfo{person}{Leigh~S. Martin}, \bibinfo{person}{Steven Martin}, \bibinfo{person}{Orion Martin}, \bibinfo{person}{Cameron Maxfield}, \bibinfo{person}{Jarrod~R. McClean}, \bibinfo{person}{Matt McEwen}, \bibinfo{person}{Seneca Meeks}, \bibinfo{person}{Anthony Megrant}, \bibinfo{person}{Xiao Mi}, \bibinfo{person}{Kevin~C. Miao}, \bibinfo{person}{Amanda Mieszala}, \bibinfo{person}{Reza Molavi}, \bibinfo{person}{Sebastian Molina}, \bibinfo{person}{Shirin Montazeri}, \bibinfo{person}{Alexis Morvan}, \bibinfo{person}{Ramis Movassagh}, \bibinfo{person}{Wojciech Mruczkiewicz}, \bibinfo{person}{Ofer Naaman}, \bibinfo{person}{Matthew Neeley}, \bibinfo{person}{Charles Neill}, \bibinfo{person}{Ani Nersisyan}, \bibinfo{person}{Hartmut Neven}, \bibinfo{person}{Michael Newman}, \bibinfo{person}{Jiun~How Ng}, \bibinfo{person}{Anthony Nguyen}, \bibinfo{person}{Murray Nguyen}, \bibinfo{person}{Chia-Hung Ni}, \bibinfo{person}{Thomas~E. O'Brien},
  \bibinfo{person}{William~D. Oliver}, \bibinfo{person}{Alex Opremcak}, \bibinfo{person}{Kristoffer Ottosson}, \bibinfo{person}{Andre Petukhov}, \bibinfo{person}{Alex Pizzuto}, \bibinfo{person}{John Platt}, \bibinfo{person}{Rebecca Potter}, \bibinfo{person}{Orion Pritchard}, \bibinfo{person}{Leonid~P. Pryadko}, \bibinfo{person}{Chris Quintana}, \bibinfo{person}{Ganesh Ramachandran}, \bibinfo{person}{Matthew~J. Reagor}, \bibinfo{person}{David~M. Rhodes}, \bibinfo{person}{Gabrielle Roberts}, \bibinfo{person}{Eliott Rosenberg}, \bibinfo{person}{Emma Rosenfeld}, \bibinfo{person}{Pedram Roushan}, \bibinfo{person}{Nicholas~C. Rubin}, \bibinfo{person}{Negar Saei}, \bibinfo{person}{Daniel Sank}, \bibinfo{person}{Kannan Sankaragomathi}, \bibinfo{person}{Kevin~J. Satzinger}, \bibinfo{person}{Henry~F. Schurkus}, \bibinfo{person}{Christopher Schuster}, \bibinfo{person}{Andrew~W. Senior}, \bibinfo{person}{Michael~J. Shearn}, \bibinfo{person}{Aaron Shorter}, \bibinfo{person}{Noah Shutty}, \bibinfo{person}{Vladimir
  Shvarts}, \bibinfo{person}{Shraddha Singh}, \bibinfo{person}{Volodymyr Sivak}, \bibinfo{person}{Jindra Skruzny}, \bibinfo{person}{Spencer Small}, \bibinfo{person}{Vadim Smelyanskiy}, \bibinfo{person}{W.~Clarke Smith}, \bibinfo{person}{Rolando~D. Somma}, \bibinfo{person}{Sofia Springer}, \bibinfo{person}{George Sterling}, \bibinfo{person}{Doug Strain}, \bibinfo{person}{Jordan Suchard}, \bibinfo{person}{Aaron Szasz}, \bibinfo{person}{Alex Sztein}, \bibinfo{person}{Douglas Thor}, \bibinfo{person}{Alfredo Torres}, \bibinfo{person}{M.~Mert Torunbalci}, \bibinfo{person}{Abeer Vaishnav}, \bibinfo{person}{Justin Vargas}, \bibinfo{person}{Sergey Vdovichev}, \bibinfo{person}{Guifre Vidal}, \bibinfo{person}{Benjamin Villalonga}, \bibinfo{person}{Catherine~Vollgraff Heidweiller}, \bibinfo{person}{Steven Waltman}, \bibinfo{person}{Shannon~X. Wang}, \bibinfo{person}{Brayden Ware}, \bibinfo{person}{Kate Weber}, \bibinfo{person}{Theodore White}, \bibinfo{person}{Kristi Wong}, \bibinfo{person}{Bryan W.~K. Woo},
  \bibinfo{person}{Cheng Xing}, \bibinfo{person}{Z.~Jamie Yao}, \bibinfo{person}{Ping Yeh}, \bibinfo{person}{Bicheng Ying}, \bibinfo{person}{Juhwan Yoo}, \bibinfo{person}{Noureldin Yosri}, \bibinfo{person}{Grayson Young}, \bibinfo{person}{Adam Zalcman}, \bibinfo{person}{Yaxing Zhang}, \bibinfo{person}{Ningfeng Zhu}, {and} \bibinfo{person}{Nicholas Zobrist}.} \bibinfo{year}{2024}\natexlab{}.
\newblock \bibinfo{title}{Quantum error correction below the surface code threshold}.
\newblock
\newblock
\showeprint[arxiv]{2408.13687}~[quant-ph]
\urldef\tempurl%
\url{https://arxiv.org/abs/2408.13687}
\showURL{%
\tempurl}


\bibitem[Arovas et~al\mbox{.}(2022)]%
        {arovas2022hubbard}
\bibfield{author}{\bibinfo{person}{Daniel~P Arovas}, \bibinfo{person}{Erez Berg}, \bibinfo{person}{Steven~A Kivelson}, {and} \bibinfo{person}{Srinivas Raghu}.} \bibinfo{year}{2022}\natexlab{}.
\newblock \showarticletitle{The hubbard model}.
\newblock \bibinfo{journal}{\emph{Annual review of condensed matter physics}} \bibinfo{volume}{13}, \bibinfo{number}{1} (\bibinfo{year}{2022}), \bibinfo{pages}{239--274}.
\newblock


\bibitem[Arute et~al\mbox{.}(2019)]%
        {arute2019quantum}
\bibfield{author}{\bibinfo{person}{Frank Arute}, \bibinfo{person}{Kunal Arya}, \bibinfo{person}{Ryan Babbush}, \bibinfo{person}{Dave Bacon}, \bibinfo{person}{Joseph~C. Bardin}, \bibinfo{person}{Rami Barends}, \bibinfo{person}{Rupak Biswas}, \bibinfo{person}{Sergio Boixo}, \bibinfo{person}{Fernando G. S.~L. Brandao}, \bibinfo{person}{David~A. Buell}, \bibinfo{person}{Brian Burkett}, \bibinfo{person}{Yu Chen}, \bibinfo{person}{Zijun Chen}, \bibinfo{person}{Ben Chiaro}, \bibinfo{person}{Roberto Collins}, \bibinfo{person}{William Courtney}, \bibinfo{person}{Andrew Dunsworth}, \bibinfo{person}{Edward Farhi}, \bibinfo{person}{Brooks Foxen}, \bibinfo{person}{Austin Fowler}, \bibinfo{person}{Craig Gidney}, \bibinfo{person}{Marissa Giustina}, \bibinfo{person}{Rob Graff}, \bibinfo{person}{Keith Guerin}, \bibinfo{person}{Steve Habegger}, \bibinfo{person}{Matthew~P. Harrigan}, \bibinfo{person}{Michael~J. Hartmann}, \bibinfo{person}{Alan Ho}, \bibinfo{person}{Markus Hoffmann}, \bibinfo{person}{Trent Huang},
  \bibinfo{person}{Travis~S. Humble}, \bibinfo{person}{Sergei~V. Isakov}, \bibinfo{person}{Evan Jeffrey}, \bibinfo{person}{Zhang Jiang}, \bibinfo{person}{Dvir Kafri}, \bibinfo{person}{Kostyantyn Kechedzhi}, \bibinfo{person}{Julian Kelly}, \bibinfo{person}{Paul~V. Klimov}, \bibinfo{person}{Sergey Knysh}, \bibinfo{person}{Alexander Korotkov}, \bibinfo{person}{Fedor Kostritsa}, \bibinfo{person}{David Landhuis}, \bibinfo{person}{Mike Lindmark}, \bibinfo{person}{Erik Lucero}, \bibinfo{person}{Dmitry Lyakh}, \bibinfo{person}{Salvatore Mandrà}, \bibinfo{person}{Jarrod~R. McClean}, \bibinfo{person}{Matthew McEwen}, \bibinfo{person}{Anthony Megrant}, \bibinfo{person}{Xiao Mi}, \bibinfo{person}{Kristel Michielsen}, \bibinfo{person}{Masoud Mohseni}, \bibinfo{person}{Josh Mutus}, \bibinfo{person}{Ofer Naaman}, \bibinfo{person}{Matthew Neeley}, \bibinfo{person}{Charles Neill}, \bibinfo{person}{Murphy~Yuezhen Niu}, \bibinfo{person}{Eric Ostby}, \bibinfo{person}{Andre Petukhov}, \bibinfo{person}{John~C. Platt},
  \bibinfo{person}{Chris Quintana}, \bibinfo{person}{Eleanor~G. Rieffel}, \bibinfo{person}{Pedram Roushan}, \bibinfo{person}{Nicholas~C. Rubin}, \bibinfo{person}{Daniel Sank}, \bibinfo{person}{Kevin~J. Satzinger}, \bibinfo{person}{Vadim Smelyanskiy}, \bibinfo{person}{Kevin~J. Sung}, \bibinfo{person}{Matthew~D. Trevithick}, \bibinfo{person}{Amit Vainsencher}, \bibinfo{person}{Benjamin Villalonga}, \bibinfo{person}{Theodore White}, \bibinfo{person}{Z.~Jamie Yao}, \bibinfo{person}{Ping Yeh}, \bibinfo{person}{Adam Zalcman}, \bibinfo{person}{Hartmut Neven}, {and} \bibinfo{person}{John~M. Martinis}.} \bibinfo{year}{2019}\natexlab{}.
\newblock \showarticletitle{Quantum supremacy using a programmable superconducting processor}.
\newblock \bibinfo{journal}{\emph{Nature}} \bibinfo{volume}{574}, \bibinfo{number}{7779} (\bibinfo{date}{Oct.} \bibinfo{year}{2019}), \bibinfo{pages}{505–510}.
\newblock
\showISSN{1476-4687}
\urldef\tempurl%
\url{https://doi.org/10.1038/s41586-019-1666-5}
\showDOI{\tempurl}


\bibitem[Auger et~al\mbox{.}(2017)]%
        {auger2017fault}
\bibfield{author}{\bibinfo{person}{James~M Auger}, \bibinfo{person}{Hussain Anwar}, \bibinfo{person}{Mercedes Gimeno-Segovia}, \bibinfo{person}{Thomas~M Stace}, {and} \bibinfo{person}{Dan~E Browne}.} \bibinfo{year}{2017}\natexlab{}.
\newblock \showarticletitle{Fault-tolerance thresholds for the surface code with fabrication errors}.
\newblock \bibinfo{journal}{\emph{Physical Review A}} \bibinfo{volume}{96}, \bibinfo{number}{4} (\bibinfo{year}{2017}), \bibinfo{pages}{042316}.
\newblock


\bibitem[Babbush et~al\mbox{.}(2018)]%
        {babbush2018encoding}
\bibfield{author}{\bibinfo{person}{Ryan Babbush}, \bibinfo{person}{Craig Gidney}, \bibinfo{person}{Dominic~W Berry}, \bibinfo{person}{Nathan Wiebe}, \bibinfo{person}{Jarrod McClean}, \bibinfo{person}{Alexandru Paler}, \bibinfo{person}{Austin Fowler}, {and} \bibinfo{person}{Hartmut Neven}.} \bibinfo{year}{2018}\natexlab{}.
\newblock \showarticletitle{Encoding electronic spectra in quantum circuits with linear T complexity}.
\newblock \bibinfo{journal}{\emph{Physical Review X}} \bibinfo{volume}{8}, \bibinfo{number}{4} (\bibinfo{year}{2018}), \bibinfo{pages}{041015}.
\newblock


\bibitem[Beverland et~al\mbox{.}(2022)]%
        {beverland2022surface}
\bibfield{author}{\bibinfo{person}{Michael Beverland}, \bibinfo{person}{Vadym Kliuchnikov}, {and} \bibinfo{person}{Eddie Schoute}.} \bibinfo{year}{2022}\natexlab{}.
\newblock \showarticletitle{Surface code compilation via edge-disjoint paths}.
\newblock \bibinfo{journal}{\emph{PRX Quantum}} \bibinfo{volume}{3}, \bibinfo{number}{2} (\bibinfo{year}{2022}), \bibinfo{pages}{020342}.
\newblock


\bibitem[Bluvstein et~al\mbox{.}(2024)]%
        {Bluvstein2024}
\bibfield{author}{\bibinfo{person}{Dolev Bluvstein}, \bibinfo{person}{Simon~J. Evered}, \bibinfo{person}{Alexandra~A. Geim}, \bibinfo{person}{Sophie~H. Li}, \bibinfo{person}{Hengyun Zhou}, \bibinfo{person}{Tom Manovitz}, \bibinfo{person}{Sepehr Ebadi}, \bibinfo{person}{Madelyn Cain}, \bibinfo{person}{Marcin Kalinowski}, \bibinfo{person}{Dominik Hangleiter}, \bibinfo{person}{J.~Pablo Bonilla~Ataides}, \bibinfo{person}{Nishad Maskara}, \bibinfo{person}{Iris Cong}, \bibinfo{person}{Xun Gao}, \bibinfo{person}{Pedro Sales~Rodriguez}, \bibinfo{person}{Thomas Karolyshyn}, \bibinfo{person}{Giulia Semeghini}, \bibinfo{person}{Michael~J. Gullans}, \bibinfo{person}{Markus Greiner}, \bibinfo{person}{Vladan Vuleti{\'{c}}}, {and} \bibinfo{person}{Mikhail~D. Lukin}.} \bibinfo{year}{2024}\natexlab{}.
\newblock \showarticletitle{Logical quantum processor based on reconfigurable atom arrays}.
\newblock \bibinfo{journal}{\emph{Nature}} \bibinfo{volume}{626}, \bibinfo{number}{7997} (\bibinfo{date}{01 Feb} \bibinfo{year}{2024}), \bibinfo{pages}{58--65}.
\newblock
\showISSN{1476-4687}
\urldef\tempurl%
\url{https://doi.org/10.1038/s41586-023-06927-3}
\showDOI{\tempurl}


\bibitem[Bomb{\'\i}n and Martin-Delgado(2009)]%
        {bombin2009quantum}
\bibfield{author}{\bibinfo{person}{H{\'e}ctor Bomb{\'\i}n} {and} \bibinfo{person}{Miguel~Angel Martin-Delgado}.} \bibinfo{year}{2009}\natexlab{}.
\newblock \showarticletitle{Quantum measurements and gates by code deformation}.
\newblock \bibinfo{journal}{\emph{Journal of Physics A: Mathematical and Theoretical}} \bibinfo{volume}{42}, \bibinfo{number}{9} (\bibinfo{year}{2009}), \bibinfo{pages}{095302}.
\newblock


\bibitem[Bravyi et~al\mbox{.}(2022)]%
        {bravyi2022future}
\bibfield{author}{\bibinfo{person}{Sergey Bravyi}, \bibinfo{person}{Oliver Dial}, \bibinfo{person}{Jay~M Gambetta}, \bibinfo{person}{Dar{\'\i}o Gil}, {and} \bibinfo{person}{Zaira Nazario}.} \bibinfo{year}{2022}\natexlab{}.
\newblock \showarticletitle{The future of quantum computing with superconducting qubits}.
\newblock \bibinfo{journal}{\emph{Journal of Applied Physics}} \bibinfo{volume}{132}, \bibinfo{number}{16} (\bibinfo{year}{2022}).
\newblock


\bibitem[Bravyi and Kitaev(1998)]%
        {bravyi1998quantum}
\bibfield{author}{\bibinfo{person}{Sergey~B Bravyi} {and} \bibinfo{person}{A~Yu Kitaev}.} \bibinfo{year}{1998}\natexlab{}.
\newblock \showarticletitle{Quantum codes on a lattice with boundary}.
\newblock \bibinfo{journal}{\emph{arXiv preprint quant-ph/9811052}} (\bibinfo{year}{1998}).
\newblock


\bibitem[Cao et~al\mbox{.}(2019)]%
        {cao2019quantum}
\bibfield{author}{\bibinfo{person}{Yudong Cao}, \bibinfo{person}{Jonathan Romero}, \bibinfo{person}{Jonathan~P. Olson}, \bibinfo{person}{Matthias Degroote}, \bibinfo{person}{Peter~D. Johnson}, \bibinfo{person}{Mária Kieferová}, \bibinfo{person}{Ian~D. Kivlichan}, \bibinfo{person}{Tim Menke}, \bibinfo{person}{Borja Peropadre}, \bibinfo{person}{Nicolas P.~D. Sawaya}, \bibinfo{person}{Sukin Sim}, \bibinfo{person}{Libor Veis}, {and} \bibinfo{person}{Alán Aspuru-Guzik}.} \bibinfo{year}{2019}\natexlab{}.
\newblock \showarticletitle{Quantum Chemistry in the Age of Quantum Computing}.
\newblock \bibinfo{journal}{\emph{Chemical Reviews}} \bibinfo{volume}{119}, \bibinfo{number}{19} (\bibinfo{date}{Oct.} \bibinfo{year}{2019}), \bibinfo{pages}{10856–10915}.
\newblock
\showISSN{0009-2665, 1520-6890}
\urldef\tempurl%
\url{https://doi.org/10.1021/acs.chemrev.8b00803}
\showDOI{\tempurl}


\bibitem[Chuang and Nielsen(1997)]%
        {chuang1997prescription}
\bibfield{author}{\bibinfo{person}{Isaac~L Chuang} {and} \bibinfo{person}{Michael~A Nielsen}.} \bibinfo{year}{1997}\natexlab{}.
\newblock \showarticletitle{Prescription for experimental determination of the dynamics of a quantum black box}.
\newblock \bibinfo{journal}{\emph{Journal of Modern Optics}} \bibinfo{volume}{44}, \bibinfo{number}{11-12} (\bibinfo{year}{1997}), \bibinfo{pages}{2455--2467}.
\newblock


\bibitem[Dennis et~al\mbox{.}(2002)]%
        {dennis2002topological}
\bibfield{author}{\bibinfo{person}{Eric Dennis}, \bibinfo{person}{Alexei Kitaev}, \bibinfo{person}{Andrew Landahl}, {and} \bibinfo{person}{John Preskill}.} \bibinfo{year}{2002}\natexlab{}.
\newblock \showarticletitle{Topological quantum memory}.
\newblock \bibinfo{journal}{\emph{J. Math. Phys.}} \bibinfo{volume}{43}, \bibinfo{number}{9} (\bibinfo{year}{2002}), \bibinfo{pages}{4452--4505}.
\newblock


\bibitem[Dong et~al\mbox{.}(2022)]%
        {dong2022beyond}
\bibfield{author}{\bibinfo{person}{Yulong Dong}, \bibinfo{person}{Jonathan Gross}, {and} \bibinfo{person}{Murphy~Yuezhen Niu}.} \bibinfo{year}{2022}\natexlab{}.
\newblock \showarticletitle{Beyond heisenberg limit quantum metrology through quantum signal processing}.
\newblock \bibinfo{journal}{\emph{arXiv preprint arXiv:2209.11207}} (\bibinfo{year}{2022}).
\newblock


\bibitem[Edman(2024)]%
        {edman2024hardware}
\bibfield{author}{\bibinfo{person}{Brayden~Thomas Edman}.} \bibinfo{year}{2024}\natexlab{}.
\newblock \showarticletitle{A Hardware-Focused Tour of IBM's 127-Qubit Eagle Processor}.
\newblock \bibinfo{journal}{\emph{Vanderbilt Undergraduate Research Journal}} \bibinfo{volume}{14}, \bibinfo{number}{1} (\bibinfo{year}{2024}).
\newblock


\bibitem[Etxezarreta~Martinez et~al\mbox{.}(2021)]%
        {etxezarreta2021time}
\bibfield{author}{\bibinfo{person}{Josu Etxezarreta~Martinez}, \bibinfo{person}{Patricio Fuentes}, \bibinfo{person}{Pedro Crespo}, {and} \bibinfo{person}{Javier Garcia-Frias}.} \bibinfo{year}{2021}\natexlab{}.
\newblock \showarticletitle{Time-varying quantum channel models for superconducting qubits}.
\newblock \bibinfo{journal}{\emph{npj Quantum Information}} \bibinfo{volume}{7}, \bibinfo{number}{1} (\bibinfo{year}{2021}), \bibinfo{pages}{115}.
\newblock


\bibitem[Etxezarreta~Martinez et~al\mbox{.}(2023)]%
        {etxezarreta2023multiqubit}
\bibfield{author}{\bibinfo{person}{Josu Etxezarreta~Martinez}, \bibinfo{person}{Patricio Fuentes}, \bibinfo{person}{Antonio deMarti iOlius}, \bibinfo{person}{Javier Garcia-Frias}, \bibinfo{person}{Javier~Rodr{\'\i}guez Fonollosa}, {and} \bibinfo{person}{Pedro~M Crespo}.} \bibinfo{year}{2023}\natexlab{}.
\newblock \showarticletitle{Multiqubit time-varying quantum channels for NISQ-era superconducting quantum processors}.
\newblock \bibinfo{journal}{\emph{Physical Review Research}} \bibinfo{volume}{5}, \bibinfo{number}{3} (\bibinfo{year}{2023}), \bibinfo{pages}{033055}.
\newblock


\bibitem[Fowler and Gidney(2018)]%
        {fowler2018low}
\bibfield{author}{\bibinfo{person}{Austin~G Fowler} {and} \bibinfo{person}{Craig Gidney}.} \bibinfo{year}{2018}\natexlab{}.
\newblock \showarticletitle{Low overhead quantum computation using lattice surgery}.
\newblock \bibinfo{journal}{\emph{arXiv preprint arXiv:1808.06709}} (\bibinfo{year}{2018}).
\newblock


\bibitem[Fowler et~al\mbox{.}(2012a)]%
        {fowler2012surface}
\bibfield{author}{\bibinfo{person}{Austin~G Fowler}, \bibinfo{person}{Matteo Mariantoni}, \bibinfo{person}{John~M Martinis}, {and} \bibinfo{person}{Andrew~N Cleland}.} \bibinfo{year}{2012}\natexlab{a}.
\newblock \showarticletitle{Surface codes: Towards practical large-scale quantum computation}.
\newblock \bibinfo{journal}{\emph{Physical Review A}} \bibinfo{volume}{86}, \bibinfo{number}{3} (\bibinfo{year}{2012}), \bibinfo{pages}{032324}.
\newblock


\bibitem[Fowler et~al\mbox{.}(2012b)]%
        {SurfaceCode}
\bibfield{author}{\bibinfo{person}{Austin~G Fowler}, \bibinfo{person}{Matteo Mariantoni}, \bibinfo{person}{John~M Martinis}, {and} \bibinfo{person}{Andrew~N Cleland}.} \bibinfo{year}{2012}\natexlab{b}.
\newblock \showarticletitle{Surface codes: Towards practical large-scale quantum computation}.
\newblock \bibinfo{journal}{\emph{Physical Review A}} \bibinfo{volume}{86}, \bibinfo{number}{3} (\bibinfo{year}{2012}), \bibinfo{pages}{032324}.
\newblock


\bibitem[Fowler et~al\mbox{.}(2014)]%
        {fowler2014scalable}
\bibfield{author}{\bibinfo{person}{Austin~G Fowler}, \bibinfo{person}{D Sank}, \bibinfo{person}{J Kelly}, \bibinfo{person}{R Barends}, {and} \bibinfo{person}{John~M Martinis}.} \bibinfo{year}{2014}\natexlab{}.
\newblock \showarticletitle{Scalable extraction of error models from the output of error detection circuits}.
\newblock \bibinfo{journal}{\emph{arXiv preprint arXiv:1405.1454}} (\bibinfo{year}{2014}).
\newblock


\bibitem[Fujiwara(2014)]%
        {fujiwara2014instantaneous}
\bibfield{author}{\bibinfo{person}{Yuichiro Fujiwara}.} \bibinfo{year}{2014}\natexlab{}.
\newblock \showarticletitle{Instantaneous quantum channel estimation during quantum information processing}.
\newblock \bibinfo{journal}{\emph{arXiv preprint arXiv:1405.6267}} (\bibinfo{year}{2014}).
\newblock


\bibitem[Gidney(2021)]%
        {stim}
\bibfield{author}{\bibinfo{person}{Craig Gidney}.} \bibinfo{year}{2021}\natexlab{}.
\newblock \showarticletitle{Stim: a fast stabilizer circuit simulator}.
\newblock \bibinfo{journal}{\emph{Quantum}}  \bibinfo{volume}{5} (\bibinfo{year}{2021}), \bibinfo{pages}{497}.
\newblock


\bibitem[Gidney and Eker{\aa{}}(2021)]%
        {Gidney2021howtofactorbit}
\bibfield{author}{\bibinfo{person}{Craig Gidney} {and} \bibinfo{person}{Martin Eker{\aa{}}}.} \bibinfo{year}{2021}\natexlab{}.
\newblock \showarticletitle{How to factor 2048 bit {RSA} integers in 8 hours using 20 million noisy qubits}.
\newblock \bibinfo{journal}{\emph{{Quantum}}}  \bibinfo{volume}{5} (\bibinfo{date}{April} \bibinfo{year}{2021}), \bibinfo{pages}{433}.
\newblock
\showISSN{2521-327X}
\urldef\tempurl%
\url{https://doi.org/10.22331/q-2021-04-15-433}
\showDOI{\tempurl}


\bibitem[Gottesman(1996)]%
        {gottesman1996class}
\bibfield{author}{\bibinfo{person}{Daniel Gottesman}.} \bibinfo{year}{1996}\natexlab{}.
\newblock \showarticletitle{Class of quantum error-correcting codes saturating the quantum Hamming bound}.
\newblock \bibinfo{journal}{\emph{Physical Review A}} \bibinfo{volume}{54}, \bibinfo{number}{3} (\bibinfo{year}{1996}), \bibinfo{pages}{1862}.
\newblock


\bibitem[Gottesman(1998a)]%
        {gottesman1998heisenberg}
\bibfield{author}{\bibinfo{person}{Daniel Gottesman}.} \bibinfo{year}{1998}\natexlab{a}.
\newblock \bibinfo{title}{The Heisenberg Representation of Quantum Computers}.
\newblock
\newblock
\showeprint[arxiv]{quant-ph/9807006}~[quant-ph]


\bibitem[Gottesman(1998b)]%
        {gottesman1998theory}
\bibfield{author}{\bibinfo{person}{Daniel Gottesman}.} \bibinfo{year}{1998}\natexlab{b}.
\newblock \showarticletitle{Theory of fault-tolerant quantum computation}.
\newblock \bibinfo{journal}{\emph{Physical Review A}} \bibinfo{volume}{57}, \bibinfo{number}{1} (\bibinfo{year}{1998}), \bibinfo{pages}{127}.
\newblock


\bibitem[G{\"u}m{\"u}{\c{s}} et~al\mbox{.}(2023)]%
        {gumucs2023calorimetry}
\bibfield{author}{\bibinfo{person}{E. G{\"u}m{\"u}{\c{s}}}, \bibinfo{person}{D. Majidi}, \bibinfo{person}{D. Nikolić}, \bibinfo{person}{P. Raif}, \bibinfo{person}{B. Karimi}, \bibinfo{person}{J.~T. Peltonen}, \bibinfo{person}{E. Scheer}, \bibinfo{person}{J.~P. Pekola}, \bibinfo{person}{H. Courtois}, \bibinfo{person}{W. Belzig}, {and} \bibinfo{person}{C.~B. Winkelmann}.} \bibinfo{year}{2023}\natexlab{}.
\newblock \showarticletitle{Calorimetry of a phase slip in a Josephson junction}.
\newblock \bibinfo{journal}{\emph{Nature Physics}} \bibinfo{volume}{19}, \bibinfo{number}{2} (\bibinfo{year}{2023}), \bibinfo{pages}{196--200}.
\newblock


\bibitem[Gustavsson et~al\mbox{.}(2016)]%
        {gustavsson2016suppressing}
\bibfield{author}{\bibinfo{person}{Simon Gustavsson}, \bibinfo{person}{Fei Yan}, \bibinfo{person}{Gianluigi Catelani}, \bibinfo{person}{Jonas Bylander}, \bibinfo{person}{Archana Kamal}, \bibinfo{person}{Jeffrey Birenbaum}, \bibinfo{person}{David Hover}, \bibinfo{person}{Danna Rosenberg}, \bibinfo{person}{Gabriel Samach}, \bibinfo{person}{Adam~P. Sears}, \bibinfo{person}{Steven~J. Weber}, \bibinfo{person}{Jonilyn~L. Yoder}, \bibinfo{person}{John Clarke}, \bibinfo{person}{Andrew~J. Kerman}, \bibinfo{person}{Fumiki Yoshihara}, \bibinfo{person}{Yasunobu Nakamura}, \bibinfo{person}{Terry~P. Orlando}, {and} \bibinfo{person}{William~D. Oliver}.} \bibinfo{year}{2016}\natexlab{}.
\newblock \showarticletitle{Suppressing relaxation in superconducting qubits by quasiparticle pumping}.
\newblock \bibinfo{journal}{\emph{Science}} \bibinfo{volume}{354}, \bibinfo{number}{6319} (\bibinfo{year}{2016}), \bibinfo{pages}{1573--1577}.
\newblock


\bibitem[Herr et~al\mbox{.}(2017)]%
        {herr2017lattice}
\bibfield{author}{\bibinfo{person}{Daniel Herr}, \bibinfo{person}{Franco Nori}, {and} \bibinfo{person}{Simon~J Devitt}.} \bibinfo{year}{2017}\natexlab{}.
\newblock \showarticletitle{Lattice surgery translation for quantum computation}.
\newblock \bibinfo{journal}{\emph{New Journal of physics}} \bibinfo{volume}{19}, \bibinfo{number}{1} (\bibinfo{year}{2017}), \bibinfo{pages}{013034}.
\newblock


\bibitem[Higgott(2022)]%
        {higgott2022pymatching}
\bibfield{author}{\bibinfo{person}{Oscar Higgott}.} \bibinfo{year}{2022}\natexlab{}.
\newblock \showarticletitle{PyMatching: A Python package for decoding quantum codes with minimum-weight perfect matching}.
\newblock \bibinfo{journal}{\emph{ACM Transactions on Quantum Computing}} \bibinfo{volume}{3}, \bibinfo{number}{3} (\bibinfo{year}{2022}), \bibinfo{pages}{1--16}.
\newblock


\bibitem[Huang et~al\mbox{.}(2020)]%
        {huang2020superconducting}
\bibfield{author}{\bibinfo{person}{He-Liang Huang}, \bibinfo{person}{Dachao Wu}, \bibinfo{person}{Daojin Fan}, {and} \bibinfo{person}{Xiaobo Zhu}.} \bibinfo{year}{2020}\natexlab{}.
\newblock \showarticletitle{Superconducting quantum computing: a review}.
\newblock \bibinfo{journal}{\emph{Science China Information Sciences}}  \bibinfo{volume}{63} (\bibinfo{year}{2020}), \bibinfo{pages}{1--32}.
\newblock


\bibitem[Huo and Li(2017)]%
        {huo2017learning}
\bibfield{author}{\bibinfo{person}{Ming-Xia Huo} {and} \bibinfo{person}{Ying Li}.} \bibinfo{year}{2017}\natexlab{}.
\newblock \showarticletitle{Learning time-dependent noise to reduce logical errors: real time error rate estimation in quantum error correction}.
\newblock \bibinfo{journal}{\emph{New Journal of Physics}} \bibinfo{volume}{19}, \bibinfo{number}{12} (\bibinfo{year}{2017}), \bibinfo{pages}{123032}.
\newblock


\bibitem[Jurcevic et~al\mbox{.}(2021)]%
        {jurcevic2021demonstration}
\bibfield{author}{\bibinfo{person}{Petar Jurcevic}, \bibinfo{person}{Ali Javadi-Abhari}, \bibinfo{person}{Lev~S Bishop}, \bibinfo{person}{Isaac Lauer}, \bibinfo{person}{Daniela~F Bogorin}, \bibinfo{person}{Markus Brink}, \bibinfo{person}{Lauren Capelluto}, \bibinfo{person}{Oktay G{\"u}nl{\"u}k}, \bibinfo{person}{Toshinari Itoko}, \bibinfo{person}{Naoki Kanazawa}, \bibinfo{person}{Abhinav Kandala}, \bibinfo{person}{George~A Keefe}, \bibinfo{person}{Kevin Krsulich}, \bibinfo{person}{William Landers}, \bibinfo{person}{Eric~P Lewandowski}, \bibinfo{person}{Douglas~T McClure}, \bibinfo{person}{Giacomo Nannicini}, \bibinfo{person}{Adinath Narasgond}, \bibinfo{person}{Hasan~M Nayfeh}, \bibinfo{person}{Emily Pritchett}, \bibinfo{person}{Mary~Beth Rothwell}, \bibinfo{person}{Srikanth Srinivasan}, \bibinfo{person}{Neereja Sundaresan}, \bibinfo{person}{Cindy Wang}, \bibinfo{person}{Ken~X Wei}, \bibinfo{person}{Christopher~J Wood}, \bibinfo{person}{Jeng-Bang Yau}, \bibinfo{person}{Eric~J Zhang}, \bibinfo{person}{Oliver~E
  Dial}, \bibinfo{person}{Jerry~M Chow}, {and} \bibinfo{person}{Jay~M Gambetta}.} \bibinfo{year}{2021}\natexlab{}.
\newblock \showarticletitle{Demonstration of quantum volume 64 on a superconducting quantum computing system}.
\newblock \bibinfo{journal}{\emph{Quantum Science and Technology}} \bibinfo{volume}{6}, \bibinfo{number}{2} (\bibinfo{year}{2021}), \bibinfo{pages}{025020}.
\newblock


\bibitem[Kelly et~al\mbox{.}(2016)]%
        {kelly2016scalable}
\bibfield{author}{\bibinfo{person}{J. Kelly}, \bibinfo{person}{R. Barends}, \bibinfo{person}{A.~G. Fowler}, \bibinfo{person}{A. Megrant}, \bibinfo{person}{E. Jeffrey}, \bibinfo{person}{T.~C. White}, \bibinfo{person}{D. Sank}, \bibinfo{person}{J.~Y. Mutus}, \bibinfo{person}{B. Campbell}, \bibinfo{person}{Yu Chen}, \bibinfo{person}{Z. Chen}, \bibinfo{person}{B. Chiaro}, \bibinfo{person}{A. Dunsworth}, \bibinfo{person}{E. Lucero}, \bibinfo{person}{M. Neeley}, \bibinfo{person}{C. Neill}, \bibinfo{person}{P.~J.~J. O’Malley}, \bibinfo{person}{C. Quintana}, \bibinfo{person}{P. Roushan}, \bibinfo{person}{A. Vainsencher}, \bibinfo{person}{J. Wenner}, {and} \bibinfo{person}{John~M. Martinis}.} \bibinfo{year}{2016}\natexlab{}.
\newblock \showarticletitle{Scalable in situ qubit calibration during repetitive error detection}.
\newblock \bibinfo{journal}{\emph{Physical Review A}} \bibinfo{volume}{94}, \bibinfo{number}{3} (\bibinfo{year}{2016}), \bibinfo{pages}{032321}.
\newblock


\bibitem[Kelly et~al\mbox{.}(2018)]%
        {kelly2018physical}
\bibfield{author}{\bibinfo{person}{Julian Kelly}, \bibinfo{person}{Peter O'Malley}, \bibinfo{person}{Matthew Neeley}, \bibinfo{person}{Hartmut Neven}, {and} \bibinfo{person}{John~M Martinis}.} \bibinfo{year}{2018}\natexlab{}.
\newblock \showarticletitle{Physical qubit calibration on a directed acyclic graph}.
\newblock \bibinfo{journal}{\emph{arXiv preprint arXiv:1803.03226}} (\bibinfo{year}{2018}).
\newblock


\bibitem[Kim et~al\mbox{.}(2023)]%
        {kim2023design}
\bibfield{author}{\bibinfo{person}{Younghun Kim}, \bibinfo{person}{Jeongsoo Kang}, {and} \bibinfo{person}{Younghun Kwon}.} \bibinfo{year}{2023}\natexlab{}.
\newblock \showarticletitle{Design of quantum error correcting code for biased error on heavy-hexagon structure}.
\newblock \bibinfo{journal}{\emph{Quantum Information Processing}} \bibinfo{volume}{22}, \bibinfo{number}{6} (\bibinfo{year}{2023}), \bibinfo{pages}{230}.
\newblock


\bibitem[Klimov et~al\mbox{.}(2024)]%
        {klimov2024optimizing}
\bibfield{author}{\bibinfo{person}{Paul~V. Klimov}, \bibinfo{person}{Andreas Bengtsson}, \bibinfo{person}{Chris Quintana}, \bibinfo{person}{Alexandre Bourassa}, \bibinfo{person}{Sabrina Hong}, \bibinfo{person}{Andrew Dunsworth}, \bibinfo{person}{Kevin~J. Satzinger}, \bibinfo{person}{William~P. Livingston}, \bibinfo{person}{Volodymyr Sivak}, \bibinfo{person}{Murphy~Yuezhen Niu}, \bibinfo{person}{Trond~I. Andersen}, \bibinfo{person}{Yaxing Zhang}, \bibinfo{person}{Desmond Chik}, \bibinfo{person}{Zijun Chen}, \bibinfo{person}{Charles Neill}, \bibinfo{person}{Catherine Erickson}, \bibinfo{person}{Alejandro Grajales~Dau}, \bibinfo{person}{Anthony Megrant}, \bibinfo{person}{Pedram Roushan}, \bibinfo{person}{Alexander~N. Korotkov}, \bibinfo{person}{Julian Kelly}, \bibinfo{person}{Vadim Smelyanskiy}, \bibinfo{person}{Yu Chen}, {and} \bibinfo{person}{Hartmut Neven}.} \bibinfo{year}{2024}\natexlab{}.
\newblock \showarticletitle{Optimizing quantum gates towards the scale of logical qubits}.
\newblock \bibinfo{journal}{\emph{Nature Communications}} \bibinfo{volume}{15}, \bibinfo{number}{1} (\bibinfo{year}{2024}), \bibinfo{pages}{2442}.
\newblock


\bibitem[Klimov et~al\mbox{.}(2018)]%
        {klimov2018fluctuations}
\bibfield{author}{\bibinfo{person}{P.~V. Klimov}, \bibinfo{person}{J. Kelly}, \bibinfo{person}{Z. Chen}, \bibinfo{person}{M. Neeley}, \bibinfo{person}{A. Megrant}, \bibinfo{person}{B. Burkett}, \bibinfo{person}{R. Barends}, \bibinfo{person}{K. Arya}, \bibinfo{person}{B. Chiaro}, \bibinfo{person}{Yu Chen}, \bibinfo{person}{A. Dunsworth}, \bibinfo{person}{A. Fowler}, \bibinfo{person}{B. Foxen}, \bibinfo{person}{C. Gidney}, \bibinfo{person}{M. Giustina}, \bibinfo{person}{R. Graff}, \bibinfo{person}{T. Huang}, \bibinfo{person}{E. Jeffrey}, \bibinfo{person}{Erik Lucero}, \bibinfo{person}{J.~Y. Mutus}, \bibinfo{person}{O. Naaman}, \bibinfo{person}{C. Neill}, \bibinfo{person}{C. Quintana}, \bibinfo{person}{P. Roushan}, \bibinfo{person}{Daniel Sank}, \bibinfo{person}{A. Vainsencher}, \bibinfo{person}{J. Wenner}, \bibinfo{person}{T.~C. White}, \bibinfo{person}{S. Boixo}, \bibinfo{person}{R. Babbush}, \bibinfo{person}{V.~N. Smelyanskiy}, \bibinfo{person}{H. Neven}, {and} \bibinfo{person}{John~M. Martinis}.}
  \bibinfo{year}{2018}\natexlab{}.
\newblock \showarticletitle{Fluctuations of energy-relaxation times in superconducting qubits}.
\newblock \bibinfo{journal}{\emph{Physical review letters}} \bibinfo{volume}{121}, \bibinfo{number}{9} (\bibinfo{year}{2018}), \bibinfo{pages}{090502}.
\newblock


\bibitem[Klimov et~al\mbox{.}(2020)]%
        {klimov2020snake}
\bibfield{author}{\bibinfo{person}{Paul~V Klimov}, \bibinfo{person}{Julian Kelly}, \bibinfo{person}{John~M Martinis}, {and} \bibinfo{person}{Hartmut Neven}.} \bibinfo{year}{2020}\natexlab{}.
\newblock \showarticletitle{The snake optimizer for learning quantum processor control parameters}.
\newblock \bibinfo{journal}{\emph{arXiv preprint arXiv:2006.04594}} (\bibinfo{year}{2020}).
\newblock


\bibitem[Knill et~al\mbox{.}(2008)]%
        {knill2008randomized}
\bibfield{author}{\bibinfo{person}{Emanuel Knill}, \bibinfo{person}{Dietrich Leibfried}, \bibinfo{person}{Rolf Reichle}, \bibinfo{person}{Joe Britton}, \bibinfo{person}{R~Brad Blakestad}, \bibinfo{person}{John~D Jost}, \bibinfo{person}{Chris Langer}, \bibinfo{person}{Roee Ozeri}, \bibinfo{person}{Signe Seidelin}, {and} \bibinfo{person}{David~J Wineland}.} \bibinfo{year}{2008}\natexlab{}.
\newblock \showarticletitle{Randomized benchmarking of quantum gates}.
\newblock \bibinfo{journal}{\emph{Physical Review A—Atomic, Molecular, and Optical Physics}} \bibinfo{volume}{77}, \bibinfo{number}{1} (\bibinfo{year}{2008}), \bibinfo{pages}{012307}.
\newblock


\bibitem[Lee et~al\mbox{.}(2021)]%
        {femoco}
\bibfield{author}{\bibinfo{person}{Joonho Lee}, \bibinfo{person}{Dominic~W. Berry}, \bibinfo{person}{Craig Gidney}, \bibinfo{person}{William~J. Huggins}, \bibinfo{person}{Jarrod~R. McClean}, \bibinfo{person}{Nathan Wiebe}, {and} \bibinfo{person}{Ryan Babbush}.} \bibinfo{year}{2021}\natexlab{}.
\newblock \showarticletitle{Even More Efficient Quantum Computations of Chemistry Through Tensor Hypercontraction}.
\newblock \bibinfo{journal}{\emph{PRX Quantum}}  \bibinfo{volume}{2} (\bibinfo{date}{Jul} \bibinfo{year}{2021}), \bibinfo{pages}{030305}.
\newblock
Issue 3.
\urldef\tempurl%
\url{https://doi.org/10.1103/PRXQuantum.2.030305}
\showDOI{\tempurl}


\bibitem[Li et~al\mbox{.}(2024)]%
        {li2024high}
\bibfield{author}{\bibinfo{person}{Tian-Ming Li}, \bibinfo{person}{Jia-Chi Zhang}, \bibinfo{person}{Bing-Jie Chen}, \bibinfo{person}{Kaixuan Huang}, \bibinfo{person}{Hao-Tian Liu}, \bibinfo{person}{Yong-Xi Xiao}, \bibinfo{person}{Cheng-Lin Deng}, \bibinfo{person}{Gui-Han Liang}, \bibinfo{person}{Chi-Tong Chen}, \bibinfo{person}{Yu Liu}, \bibinfo{person}{Hao Li}, \bibinfo{person}{Zhen-Ting Bao}, \bibinfo{person}{Kui Zhao}, \bibinfo{person}{Yueshan Xu}, \bibinfo{person}{Li Li}, \bibinfo{person}{Yang He}, \bibinfo{person}{Zheng-He Liu}, \bibinfo{person}{Yi-Han Yu}, \bibinfo{person}{Si-Yun Zhou}, \bibinfo{person}{Yan-Jun Liu}, \bibinfo{person}{Xiaohui Song}, \bibinfo{person}{Dongning Zheng}, \bibinfo{person}{Zhong-Cheng Xiang}, \bibinfo{person}{Yun-Hao Shi}, \bibinfo{person}{Kai Xu}, {and} \bibinfo{person}{Heng Fan}.} \bibinfo{year}{2024}\natexlab{}.
\newblock \showarticletitle{High-precision pulse calibration of tunable couplers for high-fidelity two-qubit gates in superconducting quantum processors}.
\newblock \bibinfo{journal}{\emph{arXiv preprint arXiv:2410.15041}} (\bibinfo{year}{2024}).
\newblock


\bibitem[Lidar and Brun(2013)]%
        {lidar2013quantum}
\bibfield{author}{\bibinfo{person}{Daniel~A Lidar} {and} \bibinfo{person}{Todd~A Brun}.} \bibinfo{year}{2013}\natexlab{}.
\newblock \bibinfo{booktitle}{\emph{Quantum error correction}}.
\newblock \bibinfo{publisher}{Cambridge university press}.
\newblock


\bibitem[Litinski(2019)]%
        {litinski2019game}
\bibfield{author}{\bibinfo{person}{Daniel Litinski}.} \bibinfo{year}{2019}\natexlab{}.
\newblock \showarticletitle{A game of surface codes: Large-scale quantum computing with lattice surgery}.
\newblock \bibinfo{journal}{\emph{Quantum}}  \bibinfo{volume}{3} (\bibinfo{year}{2019}), \bibinfo{pages}{128}.
\newblock


\bibitem[Litinski(2023)]%
        {litinski2023compute}
\bibfield{author}{\bibinfo{person}{Daniel Litinski}.} \bibinfo{year}{2023}\natexlab{}.
\newblock \showarticletitle{How to compute a 256-bit elliptic curve private key with only 50 million Toffoli gates}.
\newblock \bibinfo{journal}{\emph{arXiv preprint arXiv:2306.08585}} (\bibinfo{year}{2023}).
\newblock


\bibitem[Liu et~al\mbox{.}(2023)]%
        {liu2023enabling}
\bibfield{author}{\bibinfo{person}{Yiding Liu}, \bibinfo{person}{Zedong Li}, \bibinfo{person}{Alan Robertson}, \bibinfo{person}{Xin Fu}, {and} \bibinfo{person}{Shuaiwen~Leon Song}.} \bibinfo{year}{2023}\natexlab{}.
\newblock \showarticletitle{Enabling efficient real-time calibration on cloud quantum machines}.
\newblock \bibinfo{journal}{\emph{IEEE Transactions on Quantum Engineering}}  \bibinfo{volume}{4} (\bibinfo{year}{2023}), \bibinfo{pages}{1--17}.
\newblock


\bibitem[Magesan et~al\mbox{.}(2011)]%
        {magesan2011scalable}
\bibfield{author}{\bibinfo{person}{Easwar Magesan}, \bibinfo{person}{Jay~M Gambetta}, {and} \bibinfo{person}{Joseph Emerson}.} \bibinfo{year}{2011}\natexlab{}.
\newblock \showarticletitle{Scalable and robust randomized benchmarking of quantum processes}.
\newblock \bibinfo{journal}{\emph{Physical review letters}} \bibinfo{volume}{106}, \bibinfo{number}{18} (\bibinfo{year}{2011}), \bibinfo{pages}{180504}.
\newblock


\bibitem[Magesan et~al\mbox{.}(2012)]%
        {magesan2012efficient}
\bibfield{author}{\bibinfo{person}{Easwar Magesan}, \bibinfo{person}{Jay~M. Gambetta}, \bibinfo{person}{B.~R. Johnson}, \bibinfo{person}{Colm~A. Ryan}, \bibinfo{person}{Jerry~M. Chow}, \bibinfo{person}{Seth~T. Merkel}, \bibinfo{person}{Marcus~P. Da~Silva}, \bibinfo{person}{George~A. Keefe}, \bibinfo{person}{Mary~B. Rothwell}, \bibinfo{person}{Thomas~A. Ohki}, \bibinfo{person}{Mark~B. Ketchen}, {and} \bibinfo{person}{M. Steffen}.} \bibinfo{year}{2012}\natexlab{}.
\newblock \showarticletitle{Efficient measurement of quantum gate error by interleaved randomized benchmarking}.
\newblock \bibinfo{journal}{\emph{Physical review letters}} \bibinfo{volume}{109}, \bibinfo{number}{8} (\bibinfo{year}{2012}), \bibinfo{pages}{080505}.
\newblock


\bibitem[Martinis(2021)]%
        {martinis2021saving}
\bibfield{author}{\bibinfo{person}{John~M Martinis}.} \bibinfo{year}{2021}\natexlab{}.
\newblock \showarticletitle{Saving superconducting quantum processors from decay and correlated errors generated by gamma and cosmic rays}.
\newblock \bibinfo{journal}{\emph{npj Quantum Information}} \bibinfo{volume}{7}, \bibinfo{number}{1} (\bibinfo{year}{2021}), \bibinfo{pages}{90}.
\newblock


\bibitem[Martinis et~al\mbox{.}(2005)]%
        {martinis2005decoherence}
\bibfield{author}{\bibinfo{person}{John~M. Martinis}, \bibinfo{person}{K.~B. Cooper}, \bibinfo{person}{R. McDermott}, \bibinfo{person}{Matthias Steffen}, \bibinfo{person}{Markus Ansmann}, \bibinfo{person}{K.~D. Osborn}, \bibinfo{person}{K. Cicak}, \bibinfo{person}{Seongshik Oh}, \bibinfo{person}{D.~P. Pappas}, \bibinfo{person}{R.~W. Simmonds}, {and} \bibinfo{person}{Clare~C. Yu}.} \bibinfo{year}{2005}\natexlab{}.
\newblock \showarticletitle{Decoherence in Josephson qubits from dielectric loss}.
\newblock \bibinfo{journal}{\emph{Physical review letters}} \bibinfo{volume}{95}, \bibinfo{number}{21} (\bibinfo{year}{2005}), \bibinfo{pages}{210503}.
\newblock


\bibitem[Martyn et~al\mbox{.}(2021)]%
        {martyn2021grand}
\bibfield{author}{\bibinfo{person}{John~M Martyn}, \bibinfo{person}{Zane~M Rossi}, \bibinfo{person}{Andrew~K Tan}, {and} \bibinfo{person}{Isaac~L Chuang}.} \bibinfo{year}{2021}\natexlab{}.
\newblock \showarticletitle{Grand unification of quantum algorithms}.
\newblock \bibinfo{journal}{\emph{PRX quantum}} \bibinfo{volume}{2}, \bibinfo{number}{4} (\bibinfo{year}{2021}), \bibinfo{pages}{040203}.
\newblock


\bibitem[McEwen et~al\mbox{.}(2022)]%
        {mcewen2022resolving}
\bibfield{author}{\bibinfo{person}{Matt McEwen}, \bibinfo{person}{Lara Faoro}, \bibinfo{person}{Kunal Arya}, \bibinfo{person}{Andrew Dunsworth}, \bibinfo{person}{Trent Huang}, \bibinfo{person}{Seon Kim}, \bibinfo{person}{Brian Burkett}, \bibinfo{person}{Austin Fowler}, \bibinfo{person}{Frank Arute}, \bibinfo{person}{Joseph~C. Bardin}, \bibinfo{person}{Andreas Bengtsson}, \bibinfo{person}{Alexander Bilmes}, \bibinfo{person}{Bob~B. Buckley}, \bibinfo{person}{Nicholas Bushnell}, \bibinfo{person}{Zijun Chen}, \bibinfo{person}{Roberto Collins}, \bibinfo{person}{Sean Demura}, \bibinfo{person}{Alan~R. Derk}, \bibinfo{person}{Catherine Erickson}, \bibinfo{person}{Marissa Giustina}, \bibinfo{person}{Sean~D. Harrington}, \bibinfo{person}{Sabrina Hong}, \bibinfo{person}{Evan Jeffrey}, \bibinfo{person}{Julian Kelly}, \bibinfo{person}{Paul~V. Klimov}, \bibinfo{person}{Fedor Kostritsa}, \bibinfo{person}{Pavel Laptev}, \bibinfo{person}{Aditya Locharla}, \bibinfo{person}{Xiao Mi}, \bibinfo{person}{Kevin~C. Miao},
  \bibinfo{person}{Shirin Montazeri}, \bibinfo{person}{Josh Mutus}, \bibinfo{person}{Ofer Naaman}, \bibinfo{person}{Matthew Neeley}, \bibinfo{person}{Charles Neill}, \bibinfo{person}{Alex Opremcak}, \bibinfo{person}{Chris Quintana}, \bibinfo{person}{Nicholas Redd}, \bibinfo{person}{Pedram Roushan}, \bibinfo{person}{Daniel Sank}, \bibinfo{person}{Kevin~J. Satzinger}, \bibinfo{person}{Vladimir Shvarts}, \bibinfo{person}{Theodore White}, \bibinfo{person}{Z.~Jamie Yao}, \bibinfo{person}{Ping Yeh}, \bibinfo{person}{Juhwan Yoo}, \bibinfo{person}{Yu Chen}, \bibinfo{person}{Vadim Smelyanskiy}, \bibinfo{person}{John~M. Martinis}, \bibinfo{person}{Hartmut Neven}, \bibinfo{person}{Anthony Megrant}, \bibinfo{person}{Lev Ioffe}, {and} \bibinfo{person}{Rami Barends}.} \bibinfo{year}{2022}\natexlab{}.
\newblock \showarticletitle{Resolving catastrophic error bursts from cosmic rays in large arrays of superconducting qubits}.
\newblock \bibinfo{journal}{\emph{Nature Physics}} \bibinfo{volume}{18}, \bibinfo{number}{1} (\bibinfo{year}{2022}), \bibinfo{pages}{107--111}.
\newblock


\bibitem[M{\"u}ller et~al\mbox{.}(2019)]%
        {muller2019towards}
\bibfield{author}{\bibinfo{person}{Clemens M{\"u}ller}, \bibinfo{person}{Jared~H Cole}, {and} \bibinfo{person}{J{\"u}rgen Lisenfeld}.} \bibinfo{year}{2019}\natexlab{}.
\newblock \showarticletitle{Towards understanding two-level-systems in amorphous solids: insights from quantum circuits}.
\newblock \bibinfo{journal}{\emph{Reports on Progress in Physics}} \bibinfo{volume}{82}, \bibinfo{number}{12} (\bibinfo{year}{2019}), \bibinfo{pages}{124501}.
\newblock


\bibitem[Murali et~al\mbox{.}(2020)]%
        {murali2020software}
\bibfield{author}{\bibinfo{person}{Prakash Murali}, \bibinfo{person}{David~C McKay}, \bibinfo{person}{Margaret Martonosi}, {and} \bibinfo{person}{Ali Javadi-Abhari}.} \bibinfo{year}{2020}\natexlab{}.
\newblock \showarticletitle{Software mitigation of crosstalk on noisy intermediate-scale quantum computers}. In \bibinfo{booktitle}{\emph{Proceedings of the Twenty-Fifth International Conference on Architectural Support for Programming Languages and Operating Systems}}. \bibinfo{pages}{1001--1016}.
\newblock


\bibitem[Nagayama et~al\mbox{.}(2017)]%
        {nagayama2017surface}
\bibfield{author}{\bibinfo{person}{Shota Nagayama}, \bibinfo{person}{Austin~G Fowler}, \bibinfo{person}{Dominic Horsman}, \bibinfo{person}{Simon~J Devitt}, {and} \bibinfo{person}{Rodney Van~Meter}.} \bibinfo{year}{2017}\natexlab{}.
\newblock \showarticletitle{Surface code error correction on a defective lattice}.
\newblock \bibinfo{journal}{\emph{New Journal of Physics}} \bibinfo{volume}{19}, \bibinfo{number}{2} (\bibinfo{year}{2017}), \bibinfo{pages}{023050}.
\newblock


\bibitem[Neill et~al\mbox{.}(2018)]%
        {neill2018blueprint}
\bibfield{author}{\bibinfo{person}{C. Neill}, \bibinfo{person}{P. Roushan}, \bibinfo{person}{K. Kechedzhi}, \bibinfo{person}{S. Boixo}, \bibinfo{person}{S.~V. Isakov}, \bibinfo{person}{V. Smelyanskiy}, \bibinfo{person}{A. Megrant}, \bibinfo{person}{B. Chiaro}, \bibinfo{person}{A. Dunsworth}, \bibinfo{person}{K. Arya}, \bibinfo{person}{R. Barends}, \bibinfo{person}{B. Burkett}, \bibinfo{person}{Y. Chen}, \bibinfo{person}{Z. Chen}, \bibinfo{person}{A. Fowler}, \bibinfo{person}{B. Foxen}, \bibinfo{person}{M. Giustina}, \bibinfo{person}{R. Graff}, \bibinfo{person}{E. Jeffrey}, \bibinfo{person}{T. Huang}, \bibinfo{person}{J. Kelly}, \bibinfo{person}{P. Klimov}, \bibinfo{person}{E. Lucero}, \bibinfo{person}{J. Mutus}, \bibinfo{person}{M. Neeley}, \bibinfo{person}{C. Quintana}, \bibinfo{person}{D. Sank}, \bibinfo{person}{A. Vainsencher}, \bibinfo{person}{J. Wenner}, \bibinfo{person}{T.~C. White}, \bibinfo{person}{H. Neven}, {and} \bibinfo{person}{J.~M. Martinis}.} \bibinfo{year}{2018}\natexlab{}.
\newblock \showarticletitle{A blueprint for demonstrating quantum supremacy with superconducting qubits}.
\newblock \bibinfo{journal}{\emph{Science}} \bibinfo{volume}{360}, \bibinfo{number}{6385} (\bibinfo{year}{2018}), \bibinfo{pages}{195--199}.
\newblock


\bibitem[Nielsen and Chuang(2010)]%
        {nielsen2010quantum}
\bibfield{author}{\bibinfo{person}{Michael~A Nielsen} {and} \bibinfo{person}{Isaac~L Chuang}.} \bibinfo{year}{2010}\natexlab{}.
\newblock \bibinfo{booktitle}{\emph{Quantum computation and quantum information}}.
\newblock \bibinfo{publisher}{Cambridge university press}.
\newblock


\bibitem[Proctor et~al\mbox{.}(2020)]%
        {proctor2020detecting}
\bibfield{author}{\bibinfo{person}{Timothy Proctor}, \bibinfo{person}{Melissa Revelle}, \bibinfo{person}{Erik Nielsen}, \bibinfo{person}{Kenneth Rudinger}, \bibinfo{person}{Daniel Lobser}, \bibinfo{person}{Peter Maunz}, \bibinfo{person}{Robin Blume-Kohout}, {and} \bibinfo{person}{Kevin Young}.} \bibinfo{year}{2020}\natexlab{}.
\newblock \showarticletitle{Detecting and tracking drift in quantum information processors}.
\newblock \bibinfo{journal}{\emph{Nature communications}} \bibinfo{volume}{11}, \bibinfo{number}{1} (\bibinfo{year}{2020}), \bibinfo{pages}{5396}.
\newblock


\bibitem[Ravi et~al\mbox{.}(2023)]%
        {ravi2023navigating}
\bibfield{author}{\bibinfo{person}{Gokul~Subramanian Ravi}, \bibinfo{person}{Kaitlin Smith}, \bibinfo{person}{Jonathan~M Baker}, \bibinfo{person}{Tejas Kannan}, \bibinfo{person}{Nathan Earnest}, \bibinfo{person}{Ali Javadi-Abhari}, \bibinfo{person}{Henry Hoffmann}, {and} \bibinfo{person}{Frederic~T Chong}.} \bibinfo{year}{2023}\natexlab{}.
\newblock \showarticletitle{Navigating the dynamic noise landscape of variational quantum algorithms with QISMET}. In \bibinfo{booktitle}{\emph{Proceedings of the 28th ACM International Conference on Architectural Support for Programming Languages and Operating Systems, Volume 2}}. \bibinfo{pages}{515--529}.
\newblock


\bibitem[Shor(1996)]%
        {shor1996fault}
\bibfield{author}{\bibinfo{person}{Peter~W Shor}.} \bibinfo{year}{1996}\natexlab{}.
\newblock \showarticletitle{Fault-tolerant quantum computation}. In \bibinfo{booktitle}{\emph{Proceedings of 37th conference on foundations of computer science}}. IEEE, \bibinfo{pages}{56--65}.
\newblock


\bibitem[Shor(1999)]%
        {shor1999polynomial}
\bibfield{author}{\bibinfo{person}{Peter~W Shor}.} \bibinfo{year}{1999}\natexlab{}.
\newblock \showarticletitle{Polynomial-time algorithms for prime factorization and discrete logarithms on a quantum computer}.
\newblock \bibinfo{journal}{\emph{SIAM review}} \bibinfo{volume}{41}, \bibinfo{number}{2} (\bibinfo{year}{1999}), \bibinfo{pages}{303--332}.
\newblock


\bibitem[Springborg and Dong(2006)]%
        {jellium}
\bibfield{author}{\bibinfo{person}{M. Springborg} {and} \bibinfo{person}{Y. Dong}.} \bibinfo{year}{2006}\natexlab{}.
\newblock \showarticletitle{Chapter 4 The Jellium Model}.
\newblock In \bibinfo{booktitle}{\emph{Metallic Chains/Chains of Metals}}, \bibfield{editor}{\bibinfo{person}{Michael Springborg} {and} \bibinfo{person}{Yi~Dong}} (Eds.). \bibinfo{series}{Handbook of Metal Physics}, Vol.~\bibinfo{volume}{1}. \bibinfo{publisher}{Elsevier}, \bibinfo{pages}{37--44}.
\newblock
\showISSN{1570-002X}
\urldef\tempurl%
\url{https://doi.org/10.1016/S1570-002X(06)01004-4}
\showDOI{\tempurl}


\bibitem[Stace and Barrett(2010)]%
        {stace2010error}
\bibfield{author}{\bibinfo{person}{Thomas~M Stace} {and} \bibinfo{person}{Sean~D Barrett}.} \bibinfo{year}{2010}\natexlab{}.
\newblock \showarticletitle{Error correction and degeneracy in surface codes suffering loss}.
\newblock \bibinfo{journal}{\emph{Physical Review A}} \bibinfo{volume}{81}, \bibinfo{number}{2} (\bibinfo{year}{2010}), \bibinfo{pages}{022317}.
\newblock


\bibitem[Stace et~al\mbox{.}(2009)]%
        {stace2009thresholds}
\bibfield{author}{\bibinfo{person}{Thomas~M Stace}, \bibinfo{person}{Sean~D Barrett}, {and} \bibinfo{person}{Andrew~C Doherty}.} \bibinfo{year}{2009}\natexlab{}.
\newblock \showarticletitle{Thresholds for topological codes in the presence of loss}.
\newblock \bibinfo{journal}{\emph{Physical review letters}} \bibinfo{volume}{102}, \bibinfo{number}{20} (\bibinfo{year}{2009}), \bibinfo{pages}{200501}.
\newblock


\bibitem[Suzuki et~al\mbox{.}(2022)]%
        {suzuki2022q3de}
\bibfield{author}{\bibinfo{person}{Yasunari Suzuki}, \bibinfo{person}{Takanori Sugiyama}, \bibinfo{person}{Tomochika Arai}, \bibinfo{person}{Wang Liao}, \bibinfo{person}{Koji Inoue}, {and} \bibinfo{person}{Teruo Tanimoto}.} \bibinfo{year}{2022}\natexlab{}.
\newblock \showarticletitle{Q3DE: A fault-tolerant quantum computer architecture for multi-bit burst errors by cosmic rays}. In \bibinfo{booktitle}{\emph{2022 55th IEEE/ACM International Symposium on Microarchitecture (MICRO)}}. IEEE, \bibinfo{pages}{1110--1125}.
\newblock


\bibitem[Tannu and Qureshi(2019)]%
        {tannu2019not}
\bibfield{author}{\bibinfo{person}{Swamit~S Tannu} {and} \bibinfo{person}{Moinuddin~K Qureshi}.} \bibinfo{year}{2019}\natexlab{}.
\newblock \showarticletitle{Not all qubits are created equal: A case for variability-aware policies for NISQ-era quantum computers}. In \bibinfo{booktitle}{\emph{Proceedings of the twenty-fourth international conference on architectural support for programming languages and operating systems}}. \bibinfo{pages}{987--999}.
\newblock


\bibitem[Tornow et~al\mbox{.}(2022)]%
        {tornow2022minimum}
\bibfield{author}{\bibinfo{person}{Caroline Tornow}, \bibinfo{person}{Naoki Kanazawa}, \bibinfo{person}{William~E Shanks}, {and} \bibinfo{person}{Daniel~J Egger}.} \bibinfo{year}{2022}\natexlab{}.
\newblock \showarticletitle{Minimum quantum run-time characterization and calibration via restless measurements with dynamic repetition rates}.
\newblock \bibinfo{journal}{\emph{Physical Review Applied}} \bibinfo{volume}{17}, \bibinfo{number}{6} (\bibinfo{year}{2022}), \bibinfo{pages}{064061}.
\newblock


\bibitem[Vuillot et~al\mbox{.}(2019)]%
        {vuillot2019code}
\bibfield{author}{\bibinfo{person}{Christophe Vuillot}, \bibinfo{person}{Lingling Lao}, \bibinfo{person}{Ben Criger}, \bibinfo{person}{Carmen~Garc{\'\i}a Almud{\'e}ver}, \bibinfo{person}{Koen Bertels}, {and} \bibinfo{person}{Barbara~M Terhal}.} \bibinfo{year}{2019}\natexlab{}.
\newblock \showarticletitle{Code deformation and lattice surgery are gauge fixing}.
\newblock \bibinfo{journal}{\emph{New Journal of Physics}} \bibinfo{volume}{21}, \bibinfo{number}{3} (\bibinfo{year}{2019}), \bibinfo{pages}{033028}.
\newblock


\bibitem[Wagner et~al\mbox{.}(2021)]%
        {wagner2021optimal}
\bibfield{author}{\bibinfo{person}{Thomas Wagner}, \bibinfo{person}{Hermann Kampermann}, \bibinfo{person}{Dagmar Bru{\ss}}, {and} \bibinfo{person}{Martin Kliesch}.} \bibinfo{year}{2021}\natexlab{}.
\newblock \showarticletitle{Optimal noise estimation from syndrome statistics of quantum codes}.
\newblock \bibinfo{journal}{\emph{Physical review research}} \bibinfo{volume}{3}, \bibinfo{number}{1} (\bibinfo{year}{2021}), \bibinfo{pages}{013292}.
\newblock


\bibitem[Wagner et~al\mbox{.}(2022)]%
        {wagner2022pauli}
\bibfield{author}{\bibinfo{person}{Thomas Wagner}, \bibinfo{person}{Hermann Kampermann}, \bibinfo{person}{Dagmar Bru{\ss}}, {and} \bibinfo{person}{Martin Kliesch}.} \bibinfo{year}{2022}\natexlab{}.
\newblock \showarticletitle{Pauli channels can be estimated from syndrome measurements in quantum error correction}.
\newblock \bibinfo{journal}{\emph{Quantum}}  \bibinfo{volume}{6} (\bibinfo{year}{2022}), \bibinfo{pages}{809}.
\newblock


\bibitem[Wagner et~al\mbox{.}(2023)]%
        {wagner2023learning}
\bibfield{author}{\bibinfo{person}{Thomas Wagner}, \bibinfo{person}{Hermann Kampermann}, \bibinfo{person}{Dagmar Bru{\ss}}, {and} \bibinfo{person}{Martin Kliesch}.} \bibinfo{year}{2023}\natexlab{}.
\newblock \showarticletitle{Learning logical quantum noise in quantum error correction}.
\newblock \bibinfo{journal}{\emph{Physical review letters}} \bibinfo{volume}{130}, \bibinfo{number}{20} (\bibinfo{year}{2023}).
\newblock


\bibitem[Wang et~al\mbox{.}(2023)]%
        {wang2023dgr}
\bibfield{author}{\bibinfo{person}{Hanrui Wang}, \bibinfo{person}{Pengyu Liu}, \bibinfo{person}{Yilian Liu}, \bibinfo{person}{Jiaqi Gu}, \bibinfo{person}{Jonathan Baker}, \bibinfo{person}{Frederic~T Chong}, {and} \bibinfo{person}{Song Han}.} \bibinfo{year}{2023}\natexlab{}.
\newblock \showarticletitle{DGR: Tackling Drifted and Correlated Noise in Quantum Error Correction via Decoding Graph Re-weighting}.
\newblock \bibinfo{journal}{\emph{arXiv preprint arXiv:2311.16214}} (\bibinfo{year}{2023}).
\newblock


\bibitem[Wittler et~al\mbox{.}(2021)]%
        {wittler2021integrated}
\bibfield{author}{\bibinfo{person}{Nicolas Wittler}, \bibinfo{person}{Federico Roy}, \bibinfo{person}{Kevin Pack}, \bibinfo{person}{Max Werninghaus}, \bibinfo{person}{Anurag~Saha Roy}, \bibinfo{person}{Daniel~J Egger}, \bibinfo{person}{Stefan Filipp}, \bibinfo{person}{Frank~K Wilhelm}, {and} \bibinfo{person}{Shai Machnes}.} \bibinfo{year}{2021}\natexlab{}.
\newblock \showarticletitle{Integrated tool set for control, calibration, and characterization of quantum devices applied to superconducting qubits}.
\newblock \bibinfo{journal}{\emph{Physical Review Applied}} \bibinfo{volume}{15}, \bibinfo{number}{3} (\bibinfo{year}{2021}), \bibinfo{pages}{034080}.
\newblock


\bibitem[Yin et~al\mbox{.}(2024)]%
        {yin2024surf}
\bibfield{author}{\bibinfo{person}{Keyi Yin}, \bibinfo{person}{Xiang Fang}, \bibinfo{person}{Travis~S Humble}, \bibinfo{person}{Ang Li}, \bibinfo{person}{Yunong Shi}, {and} \bibinfo{person}{Yufei Ding}.} \bibinfo{year}{2024}\natexlab{}.
\newblock \showarticletitle{Surf-Deformer: Mitigating dynamic defects on surface code via adaptive deformation}.
\newblock  (\bibinfo{year}{2024}).
\newblock


\bibitem[Zhang et~al\mbox{.}(2020)]%
        {zhang2020error}
\bibfield{author}{\bibinfo{person}{Shuaining Zhang}, \bibinfo{person}{Yao Lu}, \bibinfo{person}{Kuan Zhang}, \bibinfo{person}{Wentao Chen}, \bibinfo{person}{Ying Li}, \bibinfo{person}{Jing-Ning Zhang}, {and} \bibinfo{person}{Kihwan Kim}.} \bibinfo{year}{2020}\natexlab{}.
\newblock \showarticletitle{Error-mitigated quantum gates exceeding physical fidelities in a trapped-ion system}.
\newblock \bibinfo{journal}{\emph{Nature communications}} \bibinfo{volume}{11}, \bibinfo{number}{1} (\bibinfo{year}{2020}), \bibinfo{pages}{587}.
\newblock


\bibitem[Zhao et~al\mbox{.}(2022)]%
        {zhao2022realization}
\bibfield{author}{\bibinfo{person}{Youwei Zhao}, \bibinfo{person}{Yangsen Ye}, \bibinfo{person}{He-Liang Huang}, \bibinfo{person}{Yiming Zhang}, \bibinfo{person}{Dachao Wu}, \bibinfo{person}{Huijie Guan}, \bibinfo{person}{Qingling Zhu}, \bibinfo{person}{Zuolin Wei}, \bibinfo{person}{Tan He}, \bibinfo{person}{Sirui Cao}, \bibinfo{person}{Fusheng Chen}, \bibinfo{person}{Tung-Hsun Chung}, \bibinfo{person}{Hui Deng}, \bibinfo{person}{Daojin Fan}, \bibinfo{person}{Ming Gong}, \bibinfo{person}{Cheng Guo}, \bibinfo{person}{Shaojun Guo}, \bibinfo{person}{Lianchen Han}, \bibinfo{person}{Na Li}, \bibinfo{person}{Shaowei Li}, \bibinfo{person}{Yuan Li}, \bibinfo{person}{Futian Liang}, \bibinfo{person}{Jin Lin}, \bibinfo{person}{Haoran Qian}, \bibinfo{person}{Hao Rong}, \bibinfo{person}{Hong Su}, \bibinfo{person}{Lihua Sun}, \bibinfo{person}{Shiyu Wang}, \bibinfo{person}{Yulin Wu}, \bibinfo{person}{Yu Xu}, \bibinfo{person}{Chong Ying}, \bibinfo{person}{Jiale Yu}, \bibinfo{person}{Chen Zha}, \bibinfo{person}{Kaili Zhang},
  \bibinfo{person}{Yong-Heng Huo}, \bibinfo{person}{Chao-Yang Lu}, \bibinfo{person}{Cheng-Zhi Peng}, \bibinfo{person}{Xiaobo Zhu}, {and} \bibinfo{person}{Jian-Wei Pan}.} \bibinfo{year}{2022}\natexlab{}.
\newblock \showarticletitle{Realization of an error-correcting surface code with superconducting qubits}.
\newblock \bibinfo{journal}{\emph{Physical Review Letters}} \bibinfo{volume}{129}, \bibinfo{number}{3} (\bibinfo{year}{2022}), \bibinfo{pages}{030501}.
\newblock


\end{thebibliography}

\end{document}